\documentclass[11pt]{article}
\usepackage{epsfig}
\newcommand{\la}[1]{\label{#1}}
\newcommand{\be}{\begin{equation}}
\newcommand{\ee}{\end{equation}}
\newcommand{\ba}{\begin{eqnarray}}
\newcommand{\ea}{\end{eqnarray}}
\newcommand{\rmi}[1]{{\mbox{\scriptsize #1}}}
\newcommand{\fig}{Fig.~}
\newcommand{\eq}{Eq.~}
\newcommand{\se}{Sec.~}
\newcommand{\eqs}{Eqs.~}
\newcommand{\nr}[1]{(\ref{#1})}

\newcommand{\nn}{\nonumber \\}
\newcommand{\fr}[2]{{\frac{#1}{#2}\,}}
\newcommand{\msbar}{{\overline{\mbox{\rm MS}}}}

\renewcommand{\vec}[1]{{\bf #1}}
\def\lsi{\raise0.3ex\hbox{$<$\kern-0.75em\raise-1.1ex\hbox{$\sim$}}}
\def\gsi{\raise0.3ex\hbox{$>$\kern-0.75em\raise-1.1ex\hbox{$\sim$}}}
\newcommand{\lsim}{\mathop{\lsi}}
\newcommand{\gsim}{\mathop{\gsi}}

\newcommand{\bmu}{\bar\mu}

\newcommand{\pmin}{\vec{p}_\rmi{min}}

\makeatletter \@addtoreset{equation}{section} \makeatother
\renewcommand{\theequation}{\arabic{section}.\arabic{equation}}
\makeatletter
\renewcommand\section{\@startsection {section}{1}{\z@}%
                                   {-5.5ex \@plus -1ex \@minus -.2ex}
                                   {2.3ex \@plus.2ex}%
                                   {\normalfont\large\bfseries}}
\renewcommand\subsection{\@startsection{subsection}{2}{\z@}%
                                     {-3.25ex\@plus -1ex \@minus -.2ex}%
                                     {1.5ex \@plus .2ex}%
                                     {\normalfont\normalsize\bfseries}}
\renewcommand\thesection {\@arabic\c@section}
\renewcommand\thesubsection   {\thesection.\@arabic\c@subsection}
\renewcommand{\@seccntformat}[1]{%
\csname the#1\endcsname.\hspace{1.0em}}
\makeatother

\begin{document}

\begin{titlepage}
\begin{flushright}
BI-TP 2004/05\\
CERN-PH-TH/2004-025\\
DAMTP-2004-10\\
HIP-2004-05/TH\\
hep-lat/0402021
\end{flushright}
\begin{centering}
\vfill
 
{\Large{\bf Duality and scaling in 3-dimensional scalar electrodynamics}}

\vspace{0.8cm}

K.~Kajantie$^{\rm a}$, 
M.~Laine$^{\rm b}$, 
T.~Neuhaus$^{\rm b}$, 
A.~Rajantie$^{\rm c}$, 
K.~Rummukainen$^{\rm d,e,f}$ 

\vspace{0.8cm}

{\em $^{\rm a}$%
Theoretical Physics Division, 
Department of Physical Sciences, \\ 
P.O.Box 64, FIN-00014 University of Helsinki, Finland\\}

\vspace{0.3cm}

{\em $^{\rm b}$%
Faculty of Physics, University of Bielefeld, 
D-33501 Bielefeld, Germany\\}

\vspace{0.3cm}

{\em $^{\rm c}$%
DAMTP, University of Cambridge, 
Wilberforce Road, Cambridge CB3 0WA, UK\\}

\vspace{0.3cm}

{\em $^{\rm d}$%
Department of Physics, Theory Division, CERN, CH-1211 Geneva 23,
Switzerland\\}

\vspace{0.3cm}

{\em $^{\rm e}$%
Department of Physics, University of Oulu, 
P.O.Box 3000, FIN-90014 Oulu, Finland\\}

\vspace{0.3cm}

{\em $^{\rm f}$%
Helsinki Institute of Physics,
P.O.Box 64, FIN-00014 University of Helsinki, Finland\\}

\vspace*{0.8cm}
 
\mbox{\bf Abstract}

\end{centering}

\vspace*{0.3cm}

\noindent
Three-dimensional scalar electrodynamics, with a local U(1) gauge
symmetry, is believed to be dual to a scalar theory with a global U(1)
symmetry, near the phase transition point. The conjectured duality
leads to definite predictions for the scaling exponents of the gauge
theory transition in the type II region, and allows thus to be
scrutinized empirically. We review these predictions, and carry out
numerical lattice Monte Carlo measurements to test them: a number of
exponents, characterising the two phases as well as the transition
point, are found to agree with expectations, supporting the
conjecture. We explain why some others, like the exponent
characterising the photon correlation length, appear to disagree with
expectations, unless very large system sizes and the extreme vicinity
of the transition point are considered.  Finally, we remark that in the 
type I region the duality implies an interesting quantitative relationship 
between a magnetic flux tube and a 2-dimensional non-topological soliton.
\vfill
\noindent
 

\vspace*{1cm}
 
\noindent
July 2004 

\vfill

\end{titlepage}

\setcounter{footnote}{0}

%
\section{Introduction}
\la{se:introduction}

Dualities constitute one of the very few analytic tools available for studying
non-perturbative properties of systems with many degrees of freedom. 
They have found applications in widely different physical settings, 
ranging from spin models to quantum field and string theories.
A duality transformation typically maps topological defects to
fundamental fields, and vice versa, and can therefore translate 
a non-perturbative problem to a solvable perturbative one, either 
in the same or in a different theory.

In some spin models, the duality transformation can be carried 
out explicitly, mapping the partition function of one theory to
that of another
(see, e.g., Refs.~\cite{Kramers:1941kn}--\cite{hk}). The duality
is therefore a mathematical identity, and one can see exactly how the 
parameters and observables of the two theories relate to each other. 
Interacting continuum field theories, on the other hand, 
are generally fairly difficult 
to treat exactly.
One reason is that they contain fluctuations 
on many different length scales, which do not decouple from each other, 
and this leads, among other things, to
ultraviolet divergences. The duality may then be approximate
rather than exact, and precisely valid only in a certain limit 
in the parameter space.

One way to limit the effect
of ultraviolet divergences is to lower the dimensionality, 
and for instance the sine-Gordon and Thirring 
models in 1+1 dimensions
have famously been shown to be dual
to each other~\cite{Coleman:bu,Mandelstam:1975hb}.
In higher dimensions, a way to control ultraviolet divergences 
is to consider supersymmetric theories, where they 
are weak or even absent.
For instance, the $\mathcal{N} = 2$ SU($N_c$) super-Yang-Mills theory
exhibits~\cite{sw} a Montonen-Olive duality~\cite{Montonen:1977sn,Osborn:tq}
between electric charges and magnetic monopoles.
Recently there has also been a great deal of interest 
in dualities between the large-$N_c$ limit of 
four-dimensional $\mathcal{N} = 4$
super-Yang-Mills theory and string
theory in five-dimensional anti-de Sitter 
background~\cite{Maldacena:1997re}.

Clearly, it would be interesting to extend 
a quantitative understanding of dualities towards 
dimensions closer to the physical 3+1, 
or to less supersymmetric theories, in order to approach, 
for instance, a consolidation of the 't Hooft-Mandelstam
dual superconductor picture of confinement in QCD
(for a recent review, see Ref.~\cite{qcd}). So far, however,
rather few examples are available in these directions. 
In this paper, we study one of them, a duality between two
{\em non-supersymmetric continuum field 
theories}, namely scalar electrodynamics (SQED) and
complex scalar field theory (SFT), in three Euclidean 
dimensions~\cite{peskin}, \cite{bmk}--\cite{son}.
The duality maps the Coulomb and Higgs phases of SQED to the broken and
symmetric phases of SFT, respectively.
The microscopic details of the theories are not dual to each other,
but as one approaches the transition point, the
duality should
become a better and better approximation at long distances. 
In principle, the duality becomes exact at the transition point
(in the so-called type II region), 
and also describes correctly the approach to the transition point.

The nature of the duality becomes more transparent if one considers
the theories in 2+1 dimensional Minkowski space. Then the duality maps
the fundamental fields of one theory to vortices of the other: the vortex 
lines can be understood as world lines of particles in the dual theory. 
One can easily make some
elementary observations that hint towards the duality. 
First, both the Coulomb phase of
SQED and the broken symmetry phase of SFT have one massless degree of
freedom, namely the photon and the Goldstone mode, 
respectively~\cite{kovner}, while in the other phase
these degrees of freedom go over into two degenerate massive
modes, in both cases.
Furthermore, both the interaction between vortices in SFT 
and the Coulomb interaction between electric charges in SQED
have the same logarithmically confining form. In contrast,
the Abrikosov-Nielsen-Olesen vortices in the type II region
of SQED and the fundamental particles of SFT with 
a positive quartic coupling, have an exponentially 
decreasing repulsive Yukawa interaction.

A full duality would, of course, be a stronger statement than simply
a counting of the degrees of freedom, and mean that 
the two theories have identical dynamics as quantum systems. 
If all the operators allowed by the symmetries were kept in the 
two theories and one knew exactly the mapping between their parameters,
the duality would predict that for any observable there is a
corresponding observable in the dual theory with the
same value. For instance, the mass spectra of the two theories should
be identical.

In practice, however, one of the theories in question
is truncated by dropping an infinite series of
high-dimen\-sional operators, and the exact
mapping between the parameters related to the operators kept is not known, 
because there is no supersymmetry to cancel radiative corrections. 
Therefore, the best way to see the duality is to calculate
universal quantities such as critical exponents, which describe the 
scaling properties of the system as the transition point is
approached. Many critical exponents of SFT are known to a high
accuracy~\cite{xy}, because the theory is in the same
universality class as the three-dimensional XY model.
The duality predicts that each of these exponents has a dual
counterpart in SQED, which should have exactly the same value.\footnote{%
 To be precise, the SQED observables we consider are sensitive
 only to the SFT exponent $\nu_\rmi{XY}$ as well as a certain 
 anomalous dimension $\eta$, while for instance the exponents 
 $\beta_\rmi{XY}, \gamma_\rmi{XY}$ related to the response 
 of SFT to an external magnetic field, 
 do not play a role in the following~\cite{peskin}.
 } 

In this paper, we 
invoke previously developed numerical techniques to study
topological defects~\cite{tension,manyvortex} and certain two-point
functions~\cite{ot,fzs} in SQED, to
measure a number of critical exponents with
lattice Monte Carlo simulations in the type II region, and compare them 
with predictions following from the duality conjecture.
The purpose is to demonstrate that these techniques yield results precise 
enough to serve as very non-trivial checks of this conjecture. 
The paper is organised as follows.
In \se\ref{se:duality}, we formulate the duality transformation
and show how it relates the critical exponents of the two theories. 
We describe our numerical simulations and present their results in 
\se\ref{se:lattice}. In \se\ref{se:Bext}, we elaborate briefly on some 
interesting qualitative manifestations of the duality conjecture in
the case of a macroscopic external magnetic field, both in the type I 
and in the type II regions of SQED, although we have not carried out any new
simulations for this situation. Finally, we discuss our findings and 
present our conclusions in \se\ref{se:concl}.

%
\section{Duality}
\la{se:duality}

%
\subsection{Basic setup}

The gauge theory under investigation, three-dimensional (3d) SQED, 
is formally defined by 
\ba
 \mathcal{L}_\rmi{SQED} & = &  
 \fr14 F_{kl}^2
 +(D_k\phi)^* (D_k \phi) 
 +m^2 \phi^*\phi 
 +\lambda \left(\phi^*\phi\right)^2
 \;, \la{Lsqed} \\
 \mathcal{Z}_\rmi{SQED} & = & 
 \int \mathcal{D} A_k \mathcal{D}\phi \mathcal{D} \phi^* 
 \exp\Bigl( -\int_x \mathcal{L}_\rmi{SQED} \Bigr)
 \;,  \la{Zsqed0}
\ea
where $F_{kl}=\partial_k A_l - \partial_l A_k$, 
$D_k=\partial_k + i e A_k$, $k,l = 1,2,3$, repeated 
indices are assumed to be summed over, and 
$\int_x \equiv \int\!{\rm d}^3 x$. This theory is 
super-renormalisable, with the only divergences appearing
in the parameter $m^2$. Unless otherwise stated, we 
assume $m^2$ to denote the bare parameter, and the theory
to be regulated, for the moment,  
with dimensional regularisation. Note that
all the arguments that follow are non-perturbative in nature, and there 
is thus no need for gauge fixing in~\eq\nr{Zsqed0}.

While the duality relation can be expressed explicitly in 
a certain deformation of SQED, namely when 
the theory is discretised and the so-called London limit
($\lambda a \to \infty$, where $a$ is the lattice spacing)
is subsequently taken~\cite{peskin}, \cite{bmk}--\cite{kleinert}, 
its status is less clear in the full continuum theory of
\eqs\nr{Lsqed}, \nr{Zsqed0}, where $\lambda$ is finite and $a\to 0$.
In fact, the phase diagram of SQED is even qualitatively different from 
that of the London limit, where the transition is
of the second order. In SQED the transition is of the first order  
if the ratio $\lambda/e^2$ is small~\cite{hlm} (the type I region), and
is believed to be of the second order for a large $\lambda/e^2$
above the Bogomolny point (the type II region). This belief is 
based, apart from the apparent vicinity of the London limit,  
on 3d renormalisation group studies~\cite{fh}--\cite{kn}, 
previous lattice simulations~\cite{dh}, \cite{b}--\cite{procs}, 
as well as other arguments~\cite{m-r}.\footnote{%
  As is well known, 
  the renormalisation group in $4-\epsilon$  dimensions, with $\epsilon\ll 1$, 
  continues to predict a first-order transition even 
  at large $\lambda/e^2$~\cite{hlm,idl}.  
  }

To now formulate the duality conjecture, 
let us choose a specific set of observables
to act as probes for the properties of the system. Defining a 
magnetic field through
\be
 B_i(x) \equiv \fr12 \epsilon_{ijk} F_{jk}(x)
 \;, \la{Bi}
\ee
we will mostly consider objects related to the two-point function
(see also Refs.~\cite{ot,fzs})
\be
 {C}_{kl}(x-y) \equiv \Bigl\langle
 B_k(x) B_l(y) \Bigr\rangle
 \;. \la{Cx}
\ee
To compute this kind of objects, let us 
generalise~\eq\nr{Zsqed0} and define
\be
 \mathcal{Z}_\rmi{SQED} [H_i] \equiv
 \int \mathcal{D} A_k \mathcal{D} \phi \mathcal{D} \phi^*
 \exp \Bigl[ 
 -\int_x \Bigl( 
 \mathcal{L}_\rmi{SQED} - H_i(x) B_i(x)
 \Bigr)
 \Bigr] \la{ZsqedH}
 \;,
\ee
where $H_i(x)$ is an arbitrary source function. The object in~\eq\nr{ZsqedH}
is in principle well defined for a complex $H_i(x)$, but in the following we
restrict ourselves
to a stripe in the complex plane where $H_i(x)$ consists of a 
real constant part (or zero-mode), 
representing a physical external magnetic 
field strength, and a purely imaginary space-dependent part, 
used as a probe to define various Green's functions. 
(If one wishes, the zero-mode could also be considered to be purely
imaginary to start with, and analytically continued to real values
only in the end, at least as long as the volume is finite.) 
The reason for this choice
is that it turns out to simplify 
the form of the corresponding SFT, without posing any restrictions 
on the physical observables that we can address.
The correlator of \eq\nr{Cx} is now obtained by taking 
the second functional derivative of 
$\ln \mathcal{Z}_\rmi{SQED} [H_i]$, and setting $H_i = 0$ afterwards.

The basic observation behind the duality is that SQED allows to define 
an exactly conserved ``charge'', the magnetic flux through some 
surface.  
This conservation law can be expressed in a local form by 
noting that, identically, 
\be
 \partial_i B_i(x) = 0 \;. \la{conserve}
\ee
Thus, there should exist some global symmetry for which
$B_i$ is the Noether current~\cite{kovner}. 
This global symmetry should be broken in 
the ``normal'' Coulomb phase where the U(1) gauge symmetry is restored, the 
massless photon representing the Goldstone boson~\cite{kovner}. 
In the ``superconducting'' Higgs
phase where the U(1) gauge symmetry is broken, on the other hand, the global 
symmetry should be restored, and the particle number of SFT, representing 
the number of Abrikosov-Nielsen-Olesen vortices in SQED, locally conserved.

If we were to stick to the Coulomb phase only, which has a single 
exactly massless degree of freedom, then the fact that an effective
description exists can be made rather 
rigorous, following the general arguments for effective
theories~\cite{weinberg}. Once we approach the phase transition point, 
however, other (non-Goldstone) degrees of freedom also become 
massless, and the situation is less clear. 

The assertion of the duality conjecture is 
that $\mathcal{Z}_\rmi{SQED} [H_i]$ equals the partition function
of a scalar field theory (SFT), 
with explicit global U(1) 
symmetry and 
a mass parameter $\tilde m^2$ which is (almost) the reverse of $m^2$, 
\be
 \tilde m^2 = c_0 - c_1 m^2 + ... 
 \;. \la{massrelation}
\ee
Here $c_1$ is dimensionless and positive, and $c_0$ is constrained
by the requirement that (with a given regularisation), 
$\tilde m^2_c = c_0 - c_1 m^2_c + ...$, where 
$\tilde m^2_c, m^2_c$ are the values of the mass parameters
at the respective transition points. The quartic coupling $\tilde \lambda$
of SFT is also constrained: 
\be
 \tilde \lambda = d_0 + d_1 \lambda + ... 
 \;, \la{lambdarelation}
\ee
where $d_1$ is dimensionless and positive, and $d_0$ is constrained
by the requirement that 
$\tilde \lambda_c = d_0 + d_1 \lambda_c + ...$, where 
$\lambda_c$ is the ``tricritical'' value separating the type I and type II
regions in SQED~\cite{kleinert2}, 
while $\tilde \lambda_c = 0$ is the value at which scalar
particles turn from attractive to repulsive  in SFT.\footnote{%
 If SFT is used for describing weakly interacting 
 atomic Bose-Einstein condensates, then
 the scalar self-coupling is usually written as 
 $\tilde \lambda = 2 \pi \hbar^2 a/m$, where $a$ is the s-wave
 scattering length and $m$ the atom mass; the transition from 
 attractive to repulsive corresponds to the s-wave scattering 
 length changing sign.}
The conjecture then reads that
\be
 \mathcal{Z}_\rmi{SQED} [H_i] = 
 \mathcal{Z}_\rmi{SFT} [H_i]
 \;, \la{conjecture}
\ee
where the form in which $H_i$ appears in the partition
function $\mathcal{Z}_\rmi{SFT}$ remains to be determined.

In order to make progress, let us
note that the symmetry underlying the duality can be considered to be  
a local one, if we assign a transformation law also to the
source field $H_i$ (cf.,\  e.g.,\ the analogous procedure in the 
case of chiral symmetry in QCD; Ref.~\cite{gl} and references therein). 
Indeed, we can allow, in general, the source term $H_i$ to change 
by a local total derivative, since this leaves 
$\mathcal{Z}_\rmi{SQED}$ invariant, 
due to~\eq\nr{conserve}, provided that the 
boundary integral emerging vanishes.
Let us inspect this issue in some more detail.

To understand the significance of the boundary term, let us place
the system in a finite box of volume $V$, and choose boundary conditions 
such that all physical fields, like $B_i(x)$, are periodic, in order
to preserve translational invariance. In accordance with our choice
for the analytic structure of $H_i(x)$, the transformation property
of $H_i$ has now to be purely imaginary,  
\be
 H_i(x) \to H'_i(x) = H_i(x) - i  \partial_i \alpha(x)
 \;, \la{sourcetrans}
\ee
where $\alpha(x)$ is an arbitrary real function, defined, 
up to a so far unfixed proportionality constant,  to be  
the generator of the local U(1) symmetry transformation.
The partition function defined in~\eq\nr{ZsqedH} has in fact a larger 
symmetry than U(1), corresponding to a complex
function $\alpha(x)$, but as we shall 
see, it is only transformations of the type in~\eq\nr{sourcetrans} which 
match the properties of the U(1) symmetry on the side of SFT. Now, since 
$\alpha(x)$ is not directly a physical field, we can in general 
consider the possibility that its boundary conditions are not strictly
periodic, but are non-periodic in some direction, by an amount 
which we call $\Delta$ for the moment. This leads, in what 
one might call ``large gauge transformations'', to
\be
 \mathcal{Z}_\rmi{SQED}[H_i'] = \mathcal{Z}_\rmi{SQED}[H_i] 
 \exp\Bigl[ i \Delta \int \! {\rm d}^2  \vec{s}\cdot \vec{B}
 \Bigr] \;,
\ee
where the integral is over that boundary of the box
at which $\alpha(x)$
is discontinuous by $\Delta$. We observe that if we choose 
$\Delta = m e$, where $m$ is an integer,
and the usual flux quantisation condition 
$e \int \! {\rm d}^2 \vec{s}\cdot \vec{B} = 2 \pi n$, 
with $n$ an integer, then the system is indeed 
fully invariant within each ``topological'' sector, 
characterised by $n$.
(The whole partition function contains a sum over 
all the sectors). 
In the following we set $m$ to its lowest non-trivial value, $m=1$, 
so that the discontinuity is $\Delta = e$.

On the SFT side, then, we assume the symmetry
transformation generated by $\alpha(x)$
to operate explicitly on the dual field variable, $\tilde \phi$: 
\be
 \tilde \phi \to \tilde \phi' = e^{i\tilde e \alpha} \tilde \phi \,\quad
 \tilde \phi^* \to \tilde \phi^*{}' = e^{-i\tilde e \alpha} \tilde \phi^* 
 \;, \la{fieldtrans} 
\ee 
where we have introduced a proportionality constant $\tilde e$. We now
see, first of all, that $\tilde e \alpha$ is defined only modulo $2\pi$.
Therefore, boundary conditions can also only be imposed modulo 
$2\pi$ for $\tilde e \alpha$, and we obtain a relation 
to the constant $\Delta$ introduced above, 
\be
 \tilde e = \frac{2\pi}{\Delta} = \frac{2 \pi}{e} \;. 
\ee
Furthermore, 
since $\mathcal{Z}_\rmi{SFT} [H_i]$ must be invariant under the 
local transformation of \eqs\nr{sourcetrans}, \nr{fieldtrans}
like $\mathcal{Z}_\rmi{SQED} [H_i]$ is, 
we can finally fix its structure: 
\ba
 \mathcal{L}_\rmi{SFT}(H_i) 
 & = &  
 \fr14 \tilde Z \tilde F_{kl}^2 
 + [(\partial_k - \tilde e H_k) \tilde \phi^*] 
   [(\partial_k + \tilde e H_k ) \tilde \phi]
 + \tilde m^2 \tilde \phi^* \tilde \phi 
 + \tilde \lambda (\tilde \phi^* \tilde \phi)^2 + ... 
 \;, \la{Lsft} \\
 \mathcal{Z}_\rmi{SFT} [H_i] 
 & = & \int \mathcal{D} \tilde \phi \mathcal{D} \tilde \phi^*
 \exp\Bigl[ - \int_x \mathcal{L}_\rmi{SFT}(H_i)  
 \Bigr] 
 \;, \la{Zsft}
\ea
where
$\tilde F_{kl} \equiv \partial_k H_l - \partial_l H_k$.

Inspecting \eq\nr{Lsft}, 
we immediately obtain a physical interpretation for the duality.
Choosing $H_k$ a real constant and coordinates so that $H_k \neq 0$
only in one direction, say $x_3$, which we may rename
to be ``imaginary time'', we note that \eqs\nr{Lsft}, \nr{Zsft}
represent just the Euclidean path integral expression 
(cf.,\ e.g., Ref.~\cite{kapusta})
for a complex SFT
with the chemical potential $\mu = \tilde e H_3$ related to the 
conserved current
\be
 \tilde j_k \equiv 2 \mathop{\mbox{Im}} 
 [\tilde \phi^* \partial_k \tilde \phi] = 
 \tilde \phi_1 \partial_k \tilde \phi_2 
 - \tilde \phi_2 \partial_k \tilde \phi_1
 \;, \la{tildejk}
\ee
where we have written $\tilde \phi$ in terms of real components as
$\tilde \phi = (\tilde \phi_1 + i \tilde \phi_2)/\sqrt{2}$. The 
corresponding total particle number 
$\int \! {\rm d}^2 \vec{s} \; \tilde j_3$
thus represents in SFT the same object as to 
which $\mu = \tilde e H_3$ couples in SQED, 
namely the integer-valued conserved flux
$(e/2\pi) \int \! {\rm d}^2 \vec{s}\cdot \vec{B}$, where ${\rm d}^2 \vec{s}$
is the spatial volume element.

As we have written SFT in \eq\nr{Lsft}, 
it only contains a few terms.
In principle, however, \eq\nr{Lsft}
should include an infinite series of 
(gauge invariant) higher order operators, starting with 
$\sim (\tilde \phi^* \tilde \phi)^3$. In the type II
region ($\tilde \lambda > 0$), they are just at most
marginal, and do not modify any of the critical exponents 
at the transition point.
In other words, their contributions
are suppressed by some power of $M/\Lambda$, where $M$ denotes
the dynamical mass scales inside the truncated
action of~\eq\nr{Lsft}, while $\Lambda \sim e^2$ is a 
``confinement'' scale related to excitations within 3d SQED
that remain non-critical at the transition point. 
In the type I region ($\tilde \lambda < 0$), 
on the other hand, the higher order operators are important. 
In fact, $\tilde \lambda < 0$ together
with a positive $(\tilde \phi^* \tilde \phi)^3$-term provides
just the usual prototype for a first order phase transition, 
as is the case in the type I region. 
In most of the discussion that follows, we consider the type II
region, and can thus ignore the higher order operators.
These arguments also implicitly
assume that the external magnetic field $\tilde e H_k$
is ``small'', i.e. at most of the same order of magnitude as 
the dynamical mass scale $M$. Note that for a constant 
$\tilde e H_k$, terms like $\tilde F_{kl}^2$ vanish.

At the end of the day, 
the external source field $H_i$ is often set to zero, 
and in that case, the Lagrangian $\mathcal{L}_\rmi{SFT}$
of \eq\nr{Lsft} simplifies to the standard one,
\be
 \mathcal{L}_\rmi{SFT}
 = 
\partial_k\tilde \phi^*
\partial_k\tilde \phi
 + \tilde m^2 \tilde \phi^* \tilde \phi 
 + \tilde \lambda (\tilde \phi^* \tilde \phi)^2 + ... 
 \;. \la{Lsft0}
\ee
The general form in Eq.~(\ref{Lsft}) is still important, however, 
because typical predictions of the duality conjecture follow 
by taking functional derivatives of the equality in~\eq\nr{conjecture}, 
and setting $H_i \to 0$ afterwards. In particular, the second
functional derivative produces for our main probe, \eq\nr{Cx}, 
\ba
 C_{kl}(x-y)
 \!\!\! & = & \!\!\! \biggl\{ 
 \frac{\delta^2\ln \mathcal{Z}_\rmi{SQED} [H_i]}{\delta H_k(x) \delta H_l(y)}
 \biggr\}_{H_i = 0} 
 \nn \!\!\! &  & \hspace*{-1.5cm} 
  = \tilde Z \Bigl( \sqcap \hspace*{-3.3mm} \sqcup 
 \delta_{kl} - \partial_k \partial_l \Bigr)
 \delta(x-y)
 + 2 \tilde e^2 \delta_{kl} \delta(x-y)
 \langle \tilde \phi^* \tilde \phi (x) \rangle 
 - \tilde e^2 
 \Bigl\langle
 \tilde j_k(x) \tilde j_l(y)  
 \Bigr\rangle
 +... \;, ~~~~~~ \la{jjrelxy}
\ea
where the expectation values on the right-hand side 
are evaluated with the 
Lagrangian of Eq.~(\ref{Lsft0}).
Both sides of the relation in \eq\nr{jjrelxy} are transverse.  
The constant $\tilde Z$ is seen to contribute to 
``contact'' terms only, $\sim \delta(x-y)$, 
but it is significant if the corresponding
susceptibility (integral over all space of the two-point 
correlation function) is considered, 
which thus is not determined by the dual theory alone.
Non-contact terms, on the other hand, are fully predicted in 
terms of the parameters of \eq\nr{Lsft0}.

To conclude, we should reiterate that in certain limits, 
for instance when SQED is replaced with a ``frozen superconductor'', 
or integer valued gauge theory, relations
of the type in~\eq\nr{jjrelxy} can be made exact, 
for any values of the parameters 
(the only one being the 
inverse temperature $\beta$ in that case)~\cite{peskin,fzs}. 
In our case, on the other hand, the relations between the parameters 
are largely open, and the parameters even get renormalised 
differently in the two theories. It is only the infrared
properties of the theory, that is correlations of the type 
in~\eq\nr{jjrelxy} at non-zero distances and close to the 
transition point, which 
can be related to each other, by suitably tuning the 
coefficients $c_i$ in~\eq\nr{massrelation}
and $d_i$ in~\eq\nr{lambdarelation}. 

%
\subsection{General structure of the photon two-point correlator}

Defining now the Fourier transform
\be
 B_k(p) \equiv \int_x e^{i p\cdot x} B_k(x) \;, 
\ee
the object we will mostly consider is the Fourier 
transform of~\eq\nr{Cx}, 
\be
 C_{kl}(p) 
 \equiv
 \frac{1}{V}
 \Bigl\langle
 B_k(-p) B_l(p)
 \Bigr\rangle
 = \int_x e^{i p\cdot(y-x)}
 \Bigl\langle
 B_k(x) B_l(y)
 \Bigr\rangle
 \equiv
 \Bigl( 
 \delta_{kl} - \frac{p_k p_l}{p^2}
 \Bigr) G(p) 
 \;, \la{jjrelp} 
\ee
where $V$ is the volume. 
We choose to measure the correlator in 
the $x_3$-direction and use a momentum transverse to that 
direction, $p\equiv \vec{p}$ 
(for instance, $\vec{p} = 2 \pi n \hat 1/ L$, 
where $\hat 1$ is the unit vector in the $x_1$-direction, 
$L$ is the linear extent of the system, 
and $n$ is an integer), 
so that
\be
 C_{33}(\vec{p}) = G(\vec{p}) \;. \la{Gdef}
\ee
The general structure of $G(\vec{p})$ is 
\be
 G(\vec{p}) = \frac{\vec{p}^2}{\vec{p}^2 + \Sigma(\vec{p})}
 \;, \la{Gstr2}
\ee
where, on the tree-level in SQED,
the self-energy is just a constant, $\Sigma(\vec{p}) = m_V^2$, 
with $m_V$ the inverse of the vector (or photon) correlation
length. More generally, we expect that close to the critical point, 
\be
 \Sigma(\vec{p}) \equiv m_\Sigma^2 +  A \, |\vec{p}|^{2 - \eta} + 
 \mathcal{O}(|\vec{p}|^\delta)
 \;, \la{Gstructure}
\ee
where $A$ is some (dimensionful) constant, $\eta$ is the 
anomalous dimension, and $\delta > 2 - \eta$.  

Taking a Fourier transform of~\eq\nr{jjrelxy}, 
$\int_x e^{i p\cdot(y-x)}(...)$, we obtain now a prediction based 
on duality for the observable of our interest, $G(\vec{p})$:  
\ba
 G(\vec{p}) 
 & = & 
 -\tilde Z \vec{p}^2 
 + 2 \tilde e^2 \langle \tilde \phi^* \tilde \phi \rangle
 - \tilde e^2 \frac{1}{V} 
 \Bigl\langle \tilde j_3(-\vec{p}) \tilde j_3(\vec{p}) \Bigr \rangle 
 + ... \;. \la{jjrel}
\ea
The non-trivial object here, 
$\Bigl\langle \tilde j_3(-\vec{p}) \tilde j_3(\vec{p}) \Bigr
\rangle$, has properties determined by SFT alone, 
and leads thus to a definite structure of $G(\vec{p})$, 
including corrections to scaling, etc. Formulating \eq\nr{jjrel} 
in such an explicit form is, as far as we know, a new result in the present
context.

In principle, \eq\nr{jjrel} could now be used to obtain 
a direct prediction for the object $G(\vec{p})$, by inserting 
the numerically measured properties of SFT  on the 
right-hand side.
In practice, this is not quite straightforward, because 
the relations of the parameters are not fixed by the conjecture. 
What can be done, however, is to use \eq\nr{jjrel} 
to obtain predictions for various critical exponents, 
since such predictions are parameter-free.
SFT is known to be in the same universality class as the
three-dimensional XY model, many critical exponents of which are known
numerically very well~\cite{xy}.
Therefore, we will use them as a
benchmark. 

%
\subsection{Critical exponents}

While SQED in the type II region
allows to define a large number of critical exponents, 
we will in this paper restrict only to a few of them, related in one
way or the other to the ``magnetic'' properties of the theory, such as
in~\eq\nr{Cx}. 
The reason is that the corresponding observables can be
measured with controllable statistical and systematic errors, thanks to 
various numerical
techniques introduced in Refs.~\cite{tension}--\cite{fzs}, and that analytic
predictions for these observables, following from the 
duality conjecture (through \eq\nr{jjrel}), are 
unambiguous. We start by considering the symmetric phase.

\paragraph{Symmetric phase: magnetic permeability.}

Magnetic permeability can be defined by considering a constant
source $H_3 \equiv  H$, and the response of the magnetic field 
strength $B \equiv V^{-1} \int_x B_3(x)$ to $H$, 
\be
 \chi \equiv  \frac{\partial B}{\partial H} 
 = \frac{1}{V} \frac{\partial^2}{\partial H^2} \ln \mathcal{Z}_\rmi{SQED}
 = \lim_{\vec{p} \to 0} G(\vec{p}) \;.
 \la{chiM}
\ee
In the free Abelian theory we would have $H = B$ and $\chi = 1$,
as can be seen by setting $\Sigma\to 0$ in~\eq\nr{Gstr2}, 
or directly by starting from the free energy $F(B) = \fr12 B^2 V$.
When interactions are taken into account, 
1-loop perturbation theory at large $m^2$ gives~\cite{prb}
\be
 \chi \approx 1 - \frac{e^2}{24\pi \sqrt{m^2(\bmu)}} + ...
 \;, \la{1lchi}
\ee 
where $m^2(\bmu)$ is the renormalised mass parameter in dimensional
regularisation (in the $\msbar$ scheme, for concreteness).
Thus, $\chi$ decreases as we approach the transition 
point $m^2(\bmu)\sim 0$.

To understand the  critical behaviour of $\chi$, note that in the broken 
phase of SFT, the correlator $\langle \tilde j_k(-p) \tilde j_l(p) \rangle$
is dominated by the massless Goldstone mode, so that 
$\langle \tilde j_k(-p) \tilde j_l(p) \rangle \propto 
\langle \tilde \phi^* \tilde \phi \rangle \; p_kp_l/p^2$, and consequently
\begin{equation}
 \lim_{\vec{p}\to 0}\langle \tilde j_3(-\vec{p})
 \tilde j_3(\vec{p}) \rangle = 0 
 \;.
\end{equation} 
Thus the duality
in~\eq\nr{jjrel} predicts that 
\be
 \chi = 2 \tilde e^2 
 \langle
 \tilde \phi^* \tilde \phi
 \rangle 
 \;. \la{pdp}
\ee
Essentially the same argument was presented recently by Son~\cite{son}: 
he considered an effective theory for the Goldstone mode alone,
$\phi(x) \equiv (1/\sqrt{2}) f \exp (i \alpha(x))$, 
and thus obtained $\chi = \tilde e^2 f^2$, 
where $f^2$ is the helicity modulus, or stiffness, or 
decay parameter, related to the Goldstone mode, and equals  
$2 \langle \tilde \phi^* \tilde \phi \rangle$ in our notation.
Note that we have implicitly assumed the use of continuum
(dimensional) regularisation in writing down the duality relation, 
and \eq\nr{pdp} should also be understood to be correspondingly
regularised. 

There are various ways to derive the critical scaling exponent of
$\langle \tilde \phi^* \tilde \phi \rangle$. 
For instance, given that the action contains 
$S_\rmi{SFT} \sim \int \! {\rm d}^3x \,
\partial_k\tilde \phi^*\partial_k\tilde \phi$, we may 
dimensionally expect~\cite{fbj} that
$\langle \tilde \phi^*\tilde \phi \rangle \sim 1/|x| \sim 1/ \xi_\rmi{SFT} \sim
|\tau|^{\nu_\rmi{XY}}$, where we have introduced 
the distance from the critical point, $\tau$, by 
\be
 \frac{m^2 - m_c^2}{e^4} \equiv \tau 
 \;.  \la{taudef}
\ee 
Other arguments leading to the same result can
be put forward~(Ref.~\cite{son} and references therein), 
and will also be met below. 
Thus, if $\chi(\tau) \sim |\tau|^{\nu_\chi}$
for $\tau \to 0^+$, we obtain~\cite{son}
\be
 \nu_\chi = \nu_\rmi{XY}
 \;, \la{nuX}
\ee
where $\nu_\rmi{XY}$ is the exponent of the correlation 
length $\xi_\rmi{SFT}$ in SFT.

To test this critical behaviour we can devise, following~\cite{ot}, 
a finite-size
scaling procedure for measuring the exponent. Close 
to the critical point, for box sizes larger than the correlation 
length $\xi$ of non-Goldstone modes, 
the system behaves as if it were almost in infinite volume, so that
($\pmin \equiv 2 \pi \hat 1/L$)
\be
 \Sigma(\pmin) = \chi^{-1}(\tau) \pmin^2 
 +\mathcal{O}(\pmin^4) 
 \;.
\ee
For box sizes smaller than $\xi$ (but still large compared with
the microscopic scales), on the other hand, the system behaves as 
if it were already at the critical point, 
\be
 \Sigma(\pmin) = A \, \pmin^{2-\eta} 
 + \mathcal{O}(\pmin^\delta)
 \;. 
 \la{Ap}
\ee
Therefore,
\be
 \pmin^\eta G^{-1}(\pmin) \sim
 \pmin^{\eta-2} \Sigma(\pmin) \sim
 \biggl\{ 
 \begin{array}{cc}
 |\tau|^{-\nu_\chi} L^{-\eta} & ,  ~~ L ~ \gsim ~ \xi   \\  
 A & , ~~ L ~ \lsim ~ \xi   
 \end{array}
 \;. 
 \la{limits}
\ee 
The function has to be continuous at 
$L\sim \xi \sim  |\tau|^{-\nu_\rmi{XY}}$; thus, 
using \eq\nr{nuX}, $\eta = 1$. 
The functional form close to the critical point should be universal, 
and according to~\eq\nr{limits} for $\eta=1$, 
only dependent on $|\tau| L^{1/\nu_\chi}$: 
\be
 \pmin G^{-1}(\pmin) 
 \sim f(|\tau| L^{1/\nu_\chi}) \;,
\ee
with $f(x) = A$, $x\ll 1$, and $f(x) \sim x^{-\nu_\chi}$, $x\gg 1$.
Subsequently at the critical point, $|\tau| \to 0$, 
it has a fixed point value, $A$. 
Moreover, if we take a derivative with respect to $\tau$
at the critical point, we obtain 
\be
 \frac{{\rm d} }{{\rm d}\tau}  \Bigl[
 \pmin G^{-1}(\pmin) \Bigr]%
 _{\tau = 0} \sim L^{1/\nu_\chi}
 \;.
 \la{fss}
\ee
\eq\nr{fss} will be used below to measure $\nu_\chi$, 
as was done in Ref.~\cite{ot} for a related model. 

\paragraph{Transition point: anomalous dimension.}

In the previous paragraph, we already found
a specific value for the anomalous dimension $\eta$, $\eta = 1$, assuming 
that the magnetic permeability scales
with the same exponent as the correlation length,
as argued in~\eq\nr{nuX}.
Let us now show that the same value for the anomalous dimension can 
be obtained without any assumption for the behaviour of $\chi(\tau)$. 
The reasoning could then be reversed, to provide yet more evidence for
the scaling of $\chi(\tau)$ according to~\eq\nr{nuX}.

The anomalous dimension related to $G(\vec{p})$ can be found
in a particularly simple way
with an argument similar to one by Son~\cite{son}. Let us 
consider $\mathcal{Z}_\rmi{SFT}[H_i]$ at the transition point. 
Factoring out the term multiplied by $\tilde Z$, the system should be
``conformally invariant'' at the critical point, 
or have no scales. For dimensional
reasons, the quadratic part of $\ln \mathcal{Z}_\rmi{SFT}$, which is 
a function of $\tilde e H_i$ only, must thus have the structure
\be
 \ln \mathcal{Z}_\rmi{SFT}[H_i] \propto \int_p
 \Bigl( \delta_{kl} - \frac{p_k p_l}{p^2} 
 \Bigr) [\tilde e H_k(-p)] [\tilde e H_l(p)] |p| 
 \;, \la{anom_form}
\ee 
where $\int_p = \int \! {\rm d}^3 p/(2 \pi)^3$. 
Using this to compute $G(\vec{p})$ through the Fourier transform
of the second functional derivative of $\ln \mathcal{Z}_\rmi{SFT}$, 
like in~\eqs\nr{jjrelxy}, \nr{jjrelp}, \nr{jjrel}, and comparing
with~\eq\nr{Gstructure} for $m_\Sigma^2 = 0$, it follows that $\eta = 1$.  

The structure of~\eq\nr{anom_form} can also be found with 
a perturbative 1-loop computation. On the side of SQED, 
the computation was carried out in Ref.~\cite{ap} in the 
Coulomb phase, 
with the result  \footnote{%
  Wilson-type renormalisation group studies in SQED 
  lead to the same functional behaviour~\cite{bfllw,ht}.}
\be
 \Sigma(\vec{p}) = \fr1{16} e^2 |\vec{p}|
 \;, ~~ \mbox{for} ~~ \vec{p}^2 \gg m^2(\bmu)
 \;.
\ee
Higher loop corrections are non-vanishing, however, 
and even diverge at the critical point $m^2(\bmu) \sim 0$, whereby there is
really no hard prediction for the critical exponent.
On the other hand, 
one could also carry out the computation on the side of SFT, 
as we just did, and possibly combine with the Wilson renormalisation 
group there: 
this line of reasoning can in principle also be used to 
understand that $\eta = 1$~(see, e.g., Ref.~\cite{ifh}). 
For a lattice study within a certain version of the dual
theory, again leading to $\eta = 1$, see Ref.~\cite{hs}.

\paragraph{Broken symmetry phase: inverse vector propagator at zero momentum.}

The quantity we will measure in the broken symmetry phase is the 
infrared limit of the inverse of the vector propagator, 
\ba
 \lim_{\vec{p} \to 0} \vec{p}^2 G^{-1}(\vec{p})
 = \lim_{\vec{p} \to 0} \Sigma(\vec{p}) \equiv m_\Sigma^2
 \;. 
 \la{mA}
\ea
Our aim is to determine the exponent
associated with $m_\Sigma^2$, 
$m_\Sigma^2 \sim |\tau|^{\gamma_\Sigma}$. To relate
$\gamma_\Sigma$ to more physical quantities, note that  
according to~\eq\nr{Gstructure}, for small $\vec{p}$ and $m_\Sigma^2$
the structure of the inverse propagator should be 
\be
 \vec{p}^2 G^{-1}(\vec{p}) \sim m_\Sigma^2 + A |\vec{p}|^{2-\eta} \;. 
\ee
Therefore the inverse of the physical vector correlation length, 
or the ``photon mass'' $m_V$, which is defined by the position of the 
singularity in $G(\vec{p})$, or zero in $\vec{p}^2 G^{-1}(\vec{p})$, 
scales as
\be
 m_V = |\vec{p}_\rmi{pole}| \sim (m_\Sigma^2)^{\frac{1}{2-\eta}}
 \sim |\tau|^{\nu_V} ~~~ \Leftrightarrow ~~~
 \nu_V = \frac{\gamma_\Sigma}{2 - \eta}
 \;.
 \la{nuV}
\ee
Thus, a determination of $\gamma_\Sigma$ combined with the known $\eta$ 
amounts to a determination of $\nu_V$. Conversely, given a prediction
for $\nu_V$, as follows in the next paragraph, we have a prediction
for $\gamma_\Sigma$~\cite{ot}, which will be tested below.

\paragraph{Broken symmetry phase: vector correlation length.}

As just mentioned, the vector correlation length is determined
by the position of the singularity in $G(\vec{p})$. According 
to~\eq\nr{jjrel}, it is determined in SFT by the 
singularity structure in the current--current correlator in the 
symmetric phase. Since the current represents a two-particle state
($\sim \tilde \phi_1 \partial \tilde \phi_2$, cf.\ \eq\nr{tildejk}), 
it is natural to expect 
that the singularity is placed at the two-particle threshold, $m_V = 2
M$, where $M$ denotes the inverse of the scalar correlation length; 
this is certainly the behaviour obtained with a 1-loop computation
in SFT. Therefore, we expect that
\be
 m_V = 2 M \sim 2 |\tau|^{\nu_\rmi{XY}}
 \;, 
 \la{mV}
\ee 
i.e., $\nu_V = \nu_\rmi{XY}$~\cite{ht,ifh}. 
Returning now back to~\eq\nr{nuV}
and inserting $\eta = 1$, 
we also find that~\cite{ot}
\be
 \gamma_\Sigma = \nu_\rmi{XY} 
 \;. \la{gammaA}
\ee

\paragraph{Broken symmetry phase: vortex tension.}

Consider finally the vortex tension. Following again an elegant argument
by Son~\cite{son}, let us consider the effect of a constant $H_3
\equiv H$ on the gauge theory side. Because of the Meissner effect, 
the system does not respond ($\mathcal{Z}_\rmi{SQED}$ does not 
change) until $H \ge H_\rmi{c1} = e T/2 \pi$, where $T$ is the 
tension of an infinitely long vortex. On the 
SFT side, on the other hand, a constant 
$\tilde e H$ corresponds to a relativistic
chemical potential $\mu$ in a (2+1)-dimensional 
theory, $\mu = \tilde e H$. Therefore, 
the system does not respond until $\mu \ge M$, where $M$ is 
the particle mass. We thus obtain $M = \tilde e H_\rmi{c1} = T$, 
i.e., that the 
vortex tension should again scale with the 
same exponent $\nu_\rmi{XY}$ as $M$ does: 
if $T \sim |\tau|^{\nu_T}$, then
\be
 \nu_{T} = \nu_\rmi{XY} 
 \;. \la{nuT}
\ee

%
\section{Simulations}
\la{se:lattice}

\begin{figure}[t]

\centerline{
\psfig{file=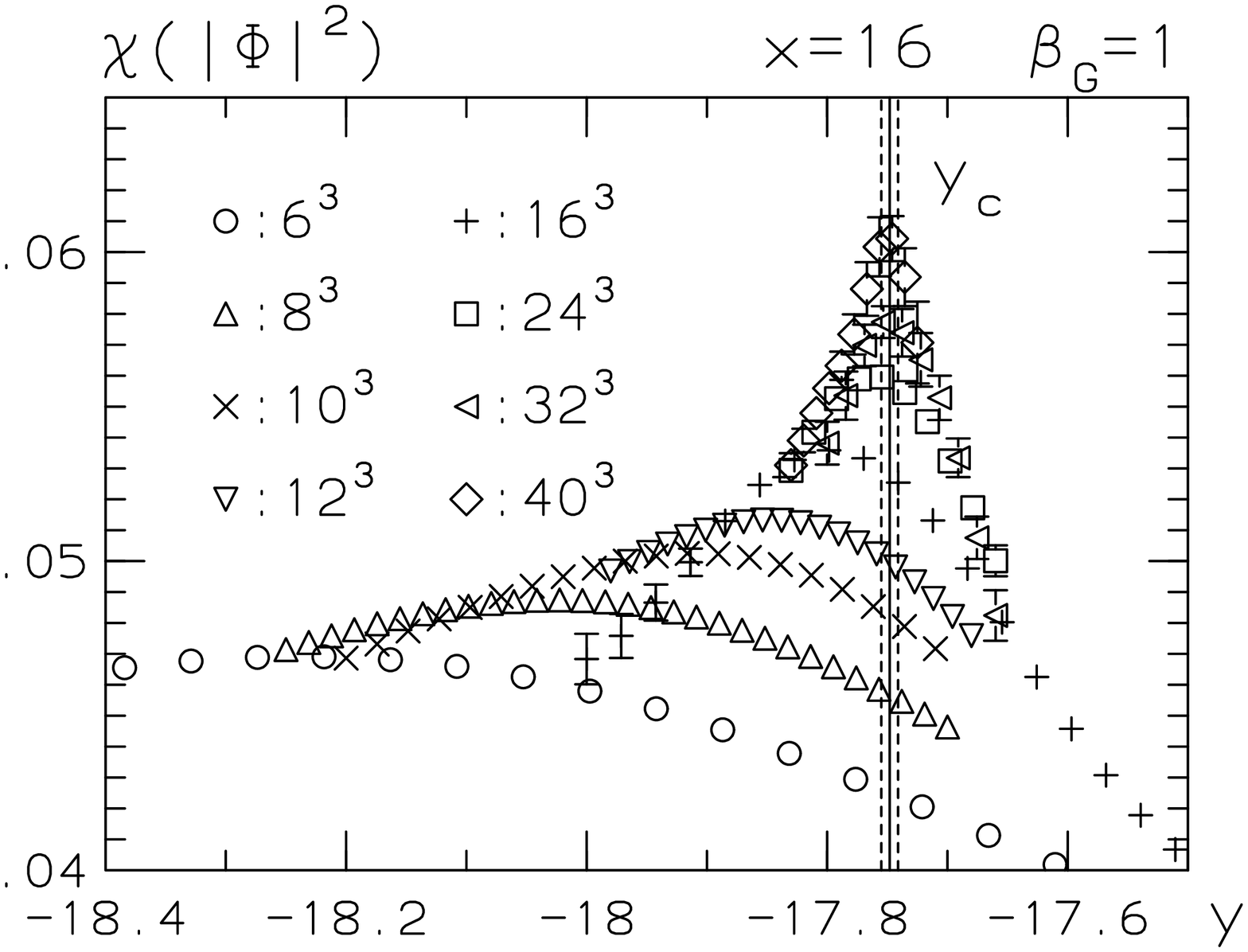,angle=270,width=7.5cm}
\psfig{file=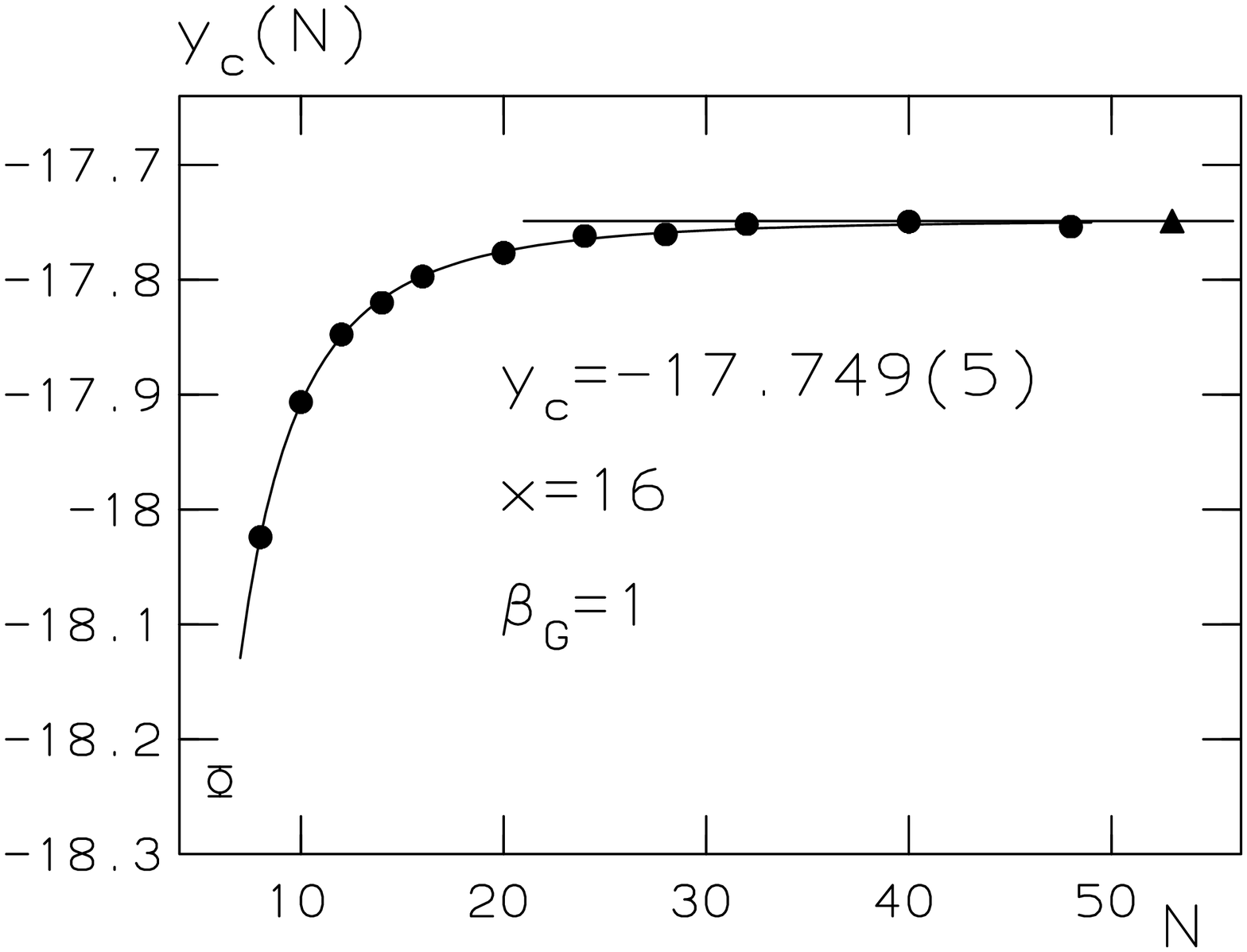,angle=270,width=7.5cm}
}

\caption[a]{ 
            Left: Examples of 
            reweighted susceptibilities $\chi(|\hat \phi|^2)$.
            Right: Determination of the 
            infinite volume critical point $y_c$
            (triangle) from the positions
            of the susceptibility maxima. 
           }

\la{fig:phi2}
\end{figure}

\begin{figure}[t]

\centerline{
\psfig{file=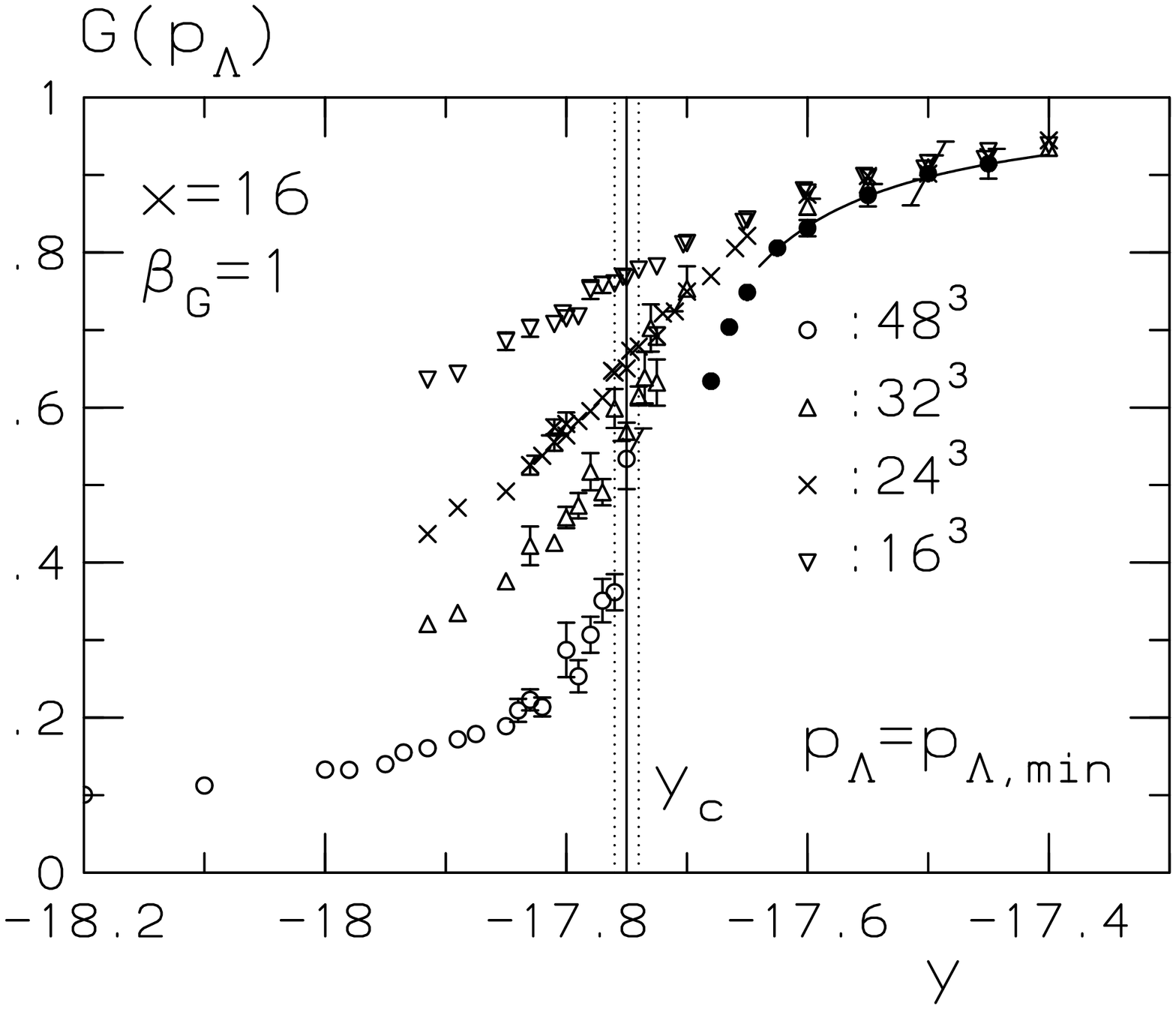,angle=270,width=7.5cm}
\psfig{file=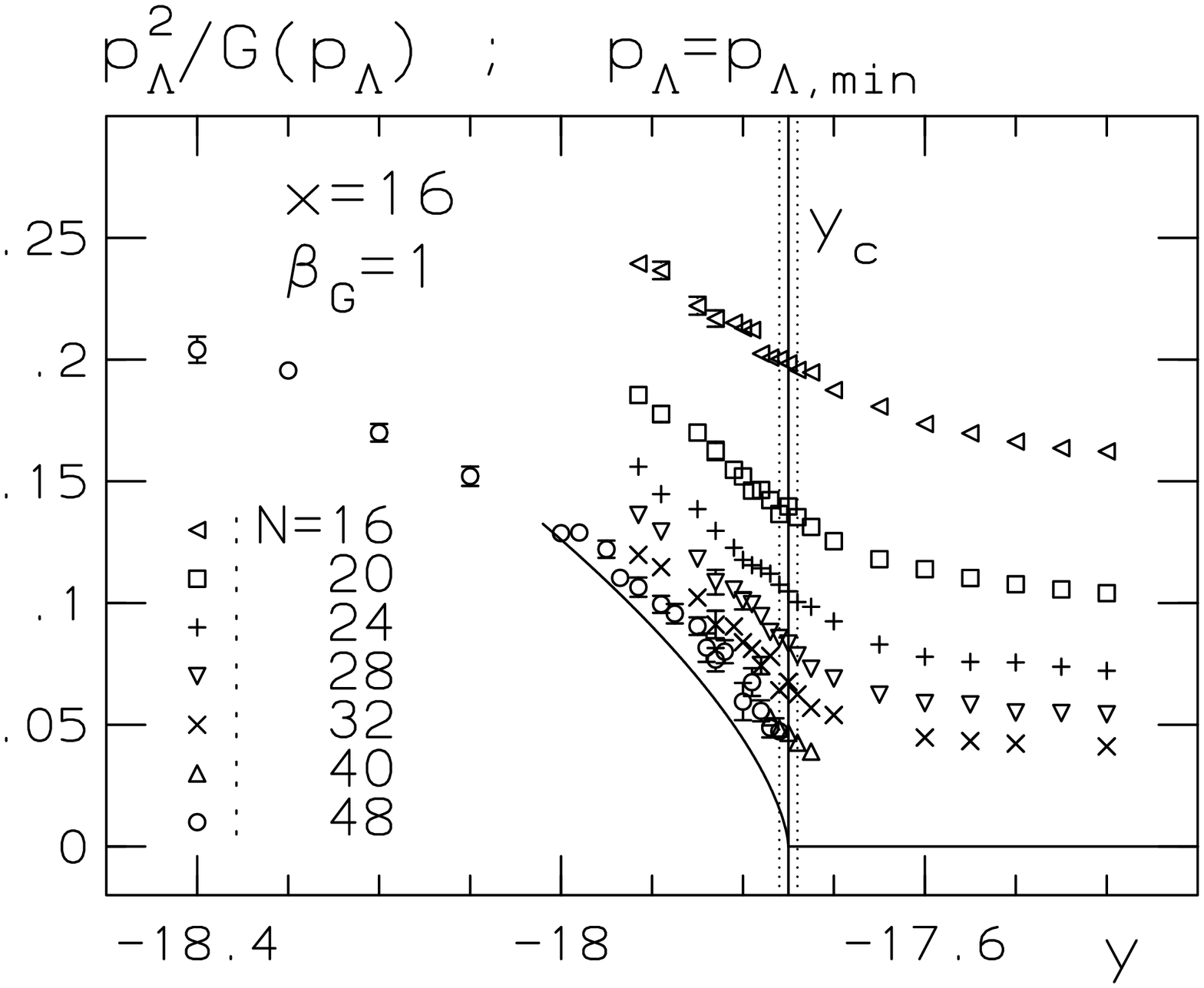,angle=270,width=7.5cm}
}

\caption[a]{ 
            Left: The behaviour of
            $G(\pmin)$ at a few representative volumes. 
            The filled circles 
            and the solid curve correspond to the infinite volume limit.
            Right: The function
            ${ \pmin^2 G^{-1}(\pmin) }$. The solid curve
            corresponds to the infinite volume limit.
           }
\la{fig:gs}
\end{figure}

\begin{figure}[t]

\centerline{
\psfig{file=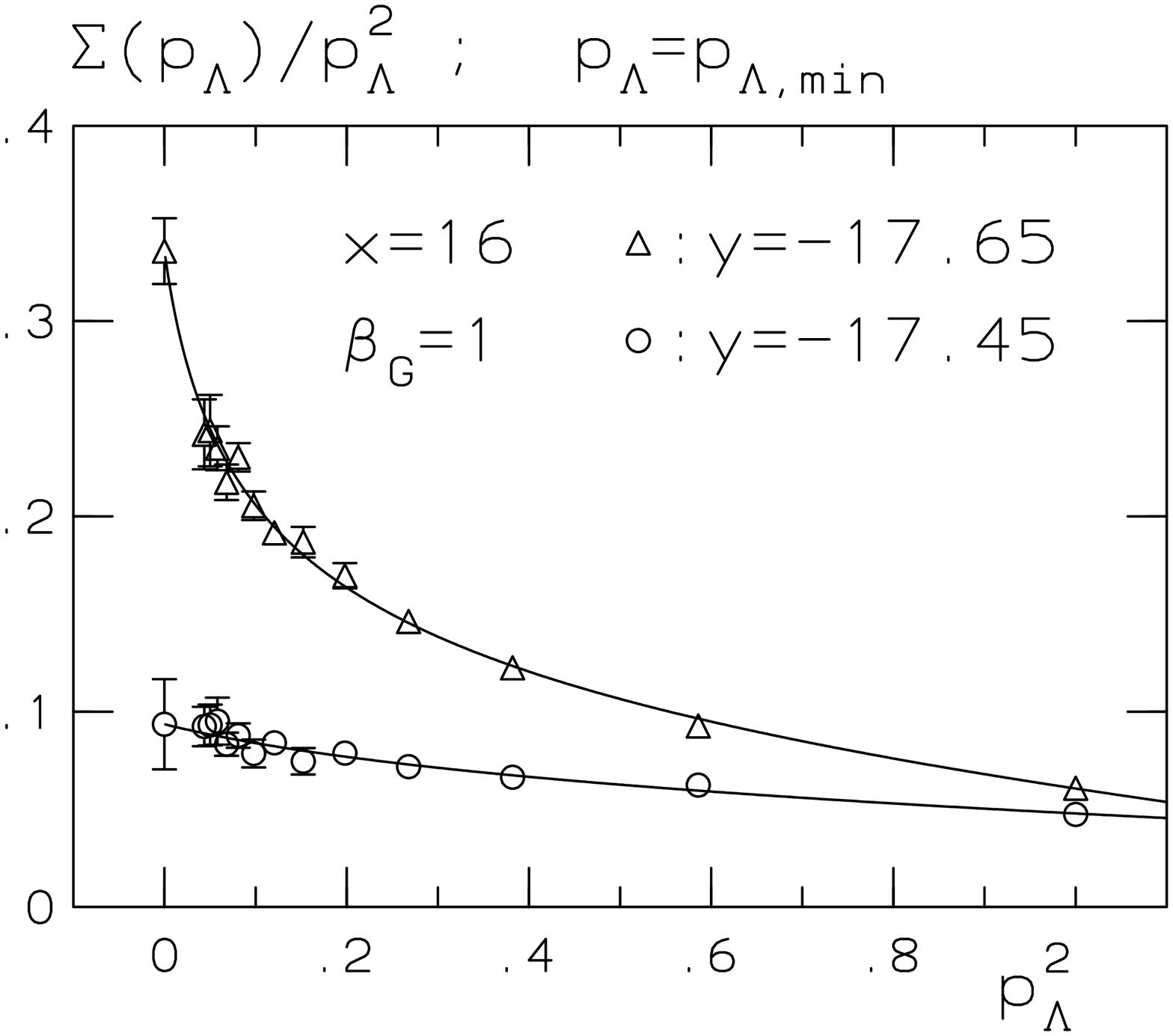,angle=270,width=7.5cm}
\psfig{file=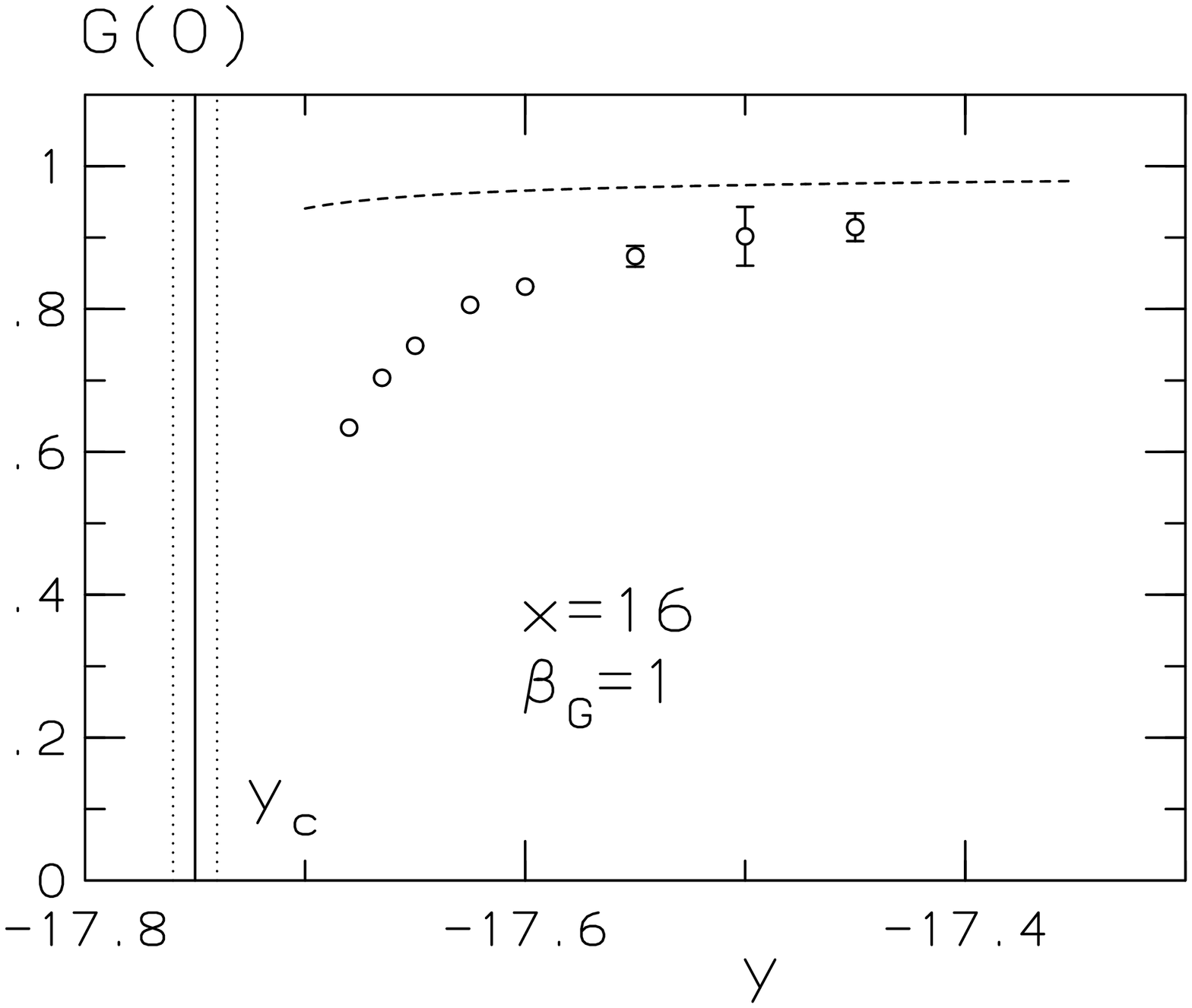,angle=270,width=7.5cm}
}

\caption[a]{ 
            Left: Examples of infinite volume extrapolations for
            $\pmin^{-2} \Sigma (\pmin)$ in the symmetric
            phase. We fit to the 1-loop perturbative expression, 
            with the mass of the scalar particle as a free parameter. 
            The extrapolation range
            increases rapidly while approaching
            the critical point. 
            Right: The corresponding behavior for
            $G(0) = \lim_{\pmin\to 0}[1+\pmin^{-2} \Sigma(\pmin)]^{-1}$. 
            The dashed line represents the perturbative prediction 
            in \eq\nr{1lchi}, with $m^2(\bmu) \to (y-y_c) e^4$.
           }

\la{fig:permn}
\end{figure}

\begin{figure}[t]

\centerline{
\psfig{file=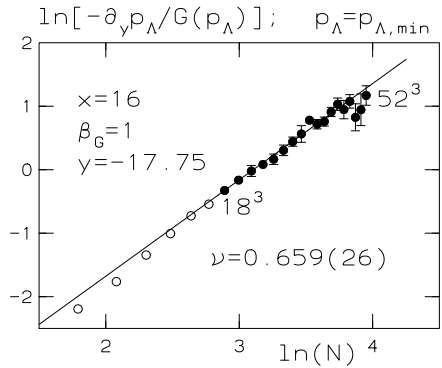,angle=270,width=7.5cm}
}

\caption[a]{ 
            The derivative
            $-\partial_y [{ \pmin G^{-1}(\pmin)}]$ 
            at the critical point, 
            as a function of the volume, and a determination
            of the exponent of magnetic permeability via \eq\nr{fss}.
           }

\la{fig:permfss}
\end{figure}

\begin{figure}[t]

\centerline{
\psfig{file=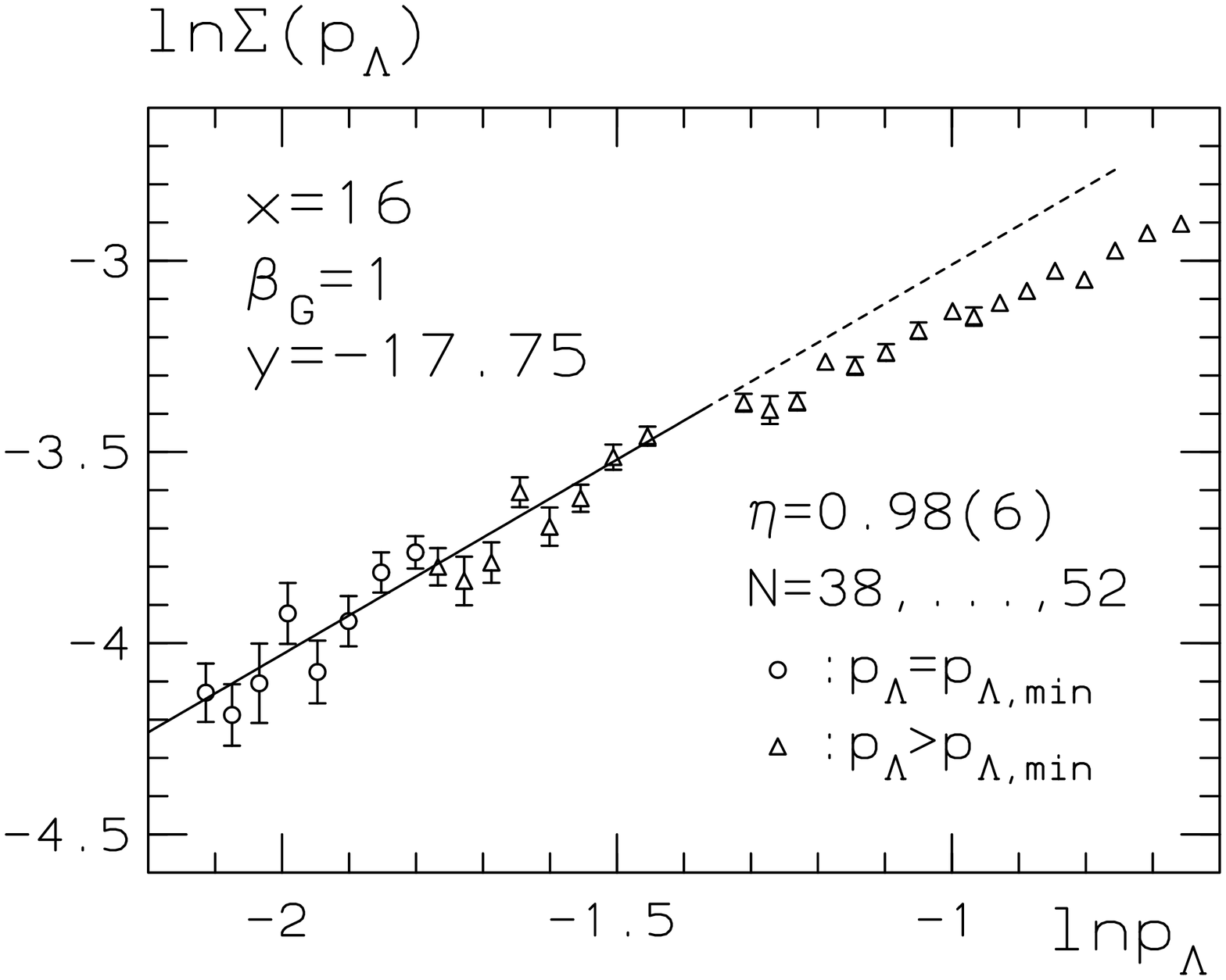,angle=270,width=7.5cm}
\psfig{file=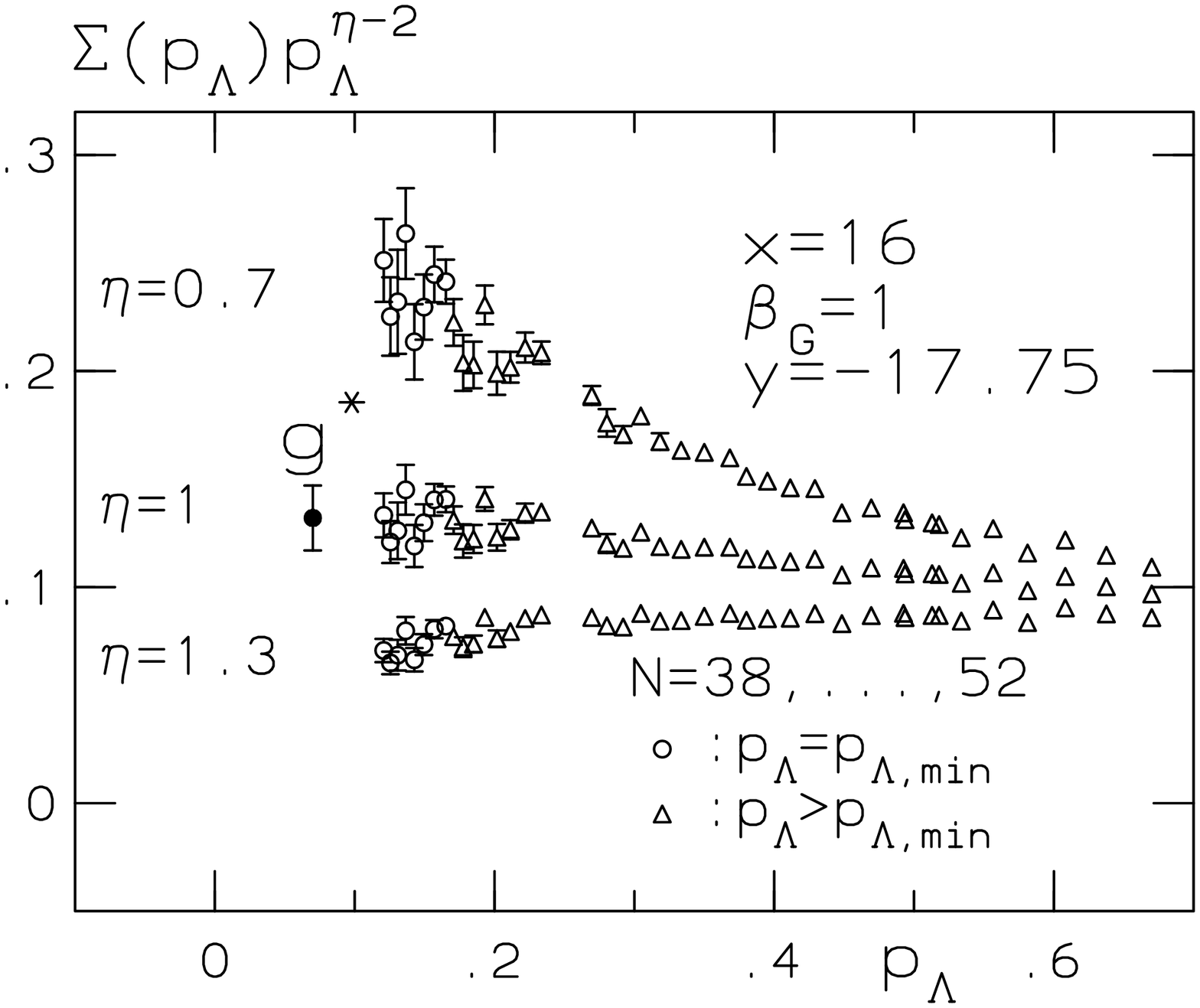,angle=270,width=7.5cm}
}
 
\caption[a]{ 
            Left: 
            The photon self-energy $\Sigma(\vec{p})$ at the critical
            point. Circles denote momenta $|\pmin|=2\pi/L$
            at different volumes; 
            triangles denote larger momenta. The solid curve is a linear
            fit determining the anomalous dimension $\eta$.
            Right: 
            An alternative determination of $\eta$
            as well as of the fixed-point coefficient $g^*$
            (cf.\ \fig\ref{fig:matching}), 
            based on~\eq\nr{limits}.
           }
\la{fig:eta}
\end{figure}

\begin{figure}[t]

\centerline{
\psfig{file=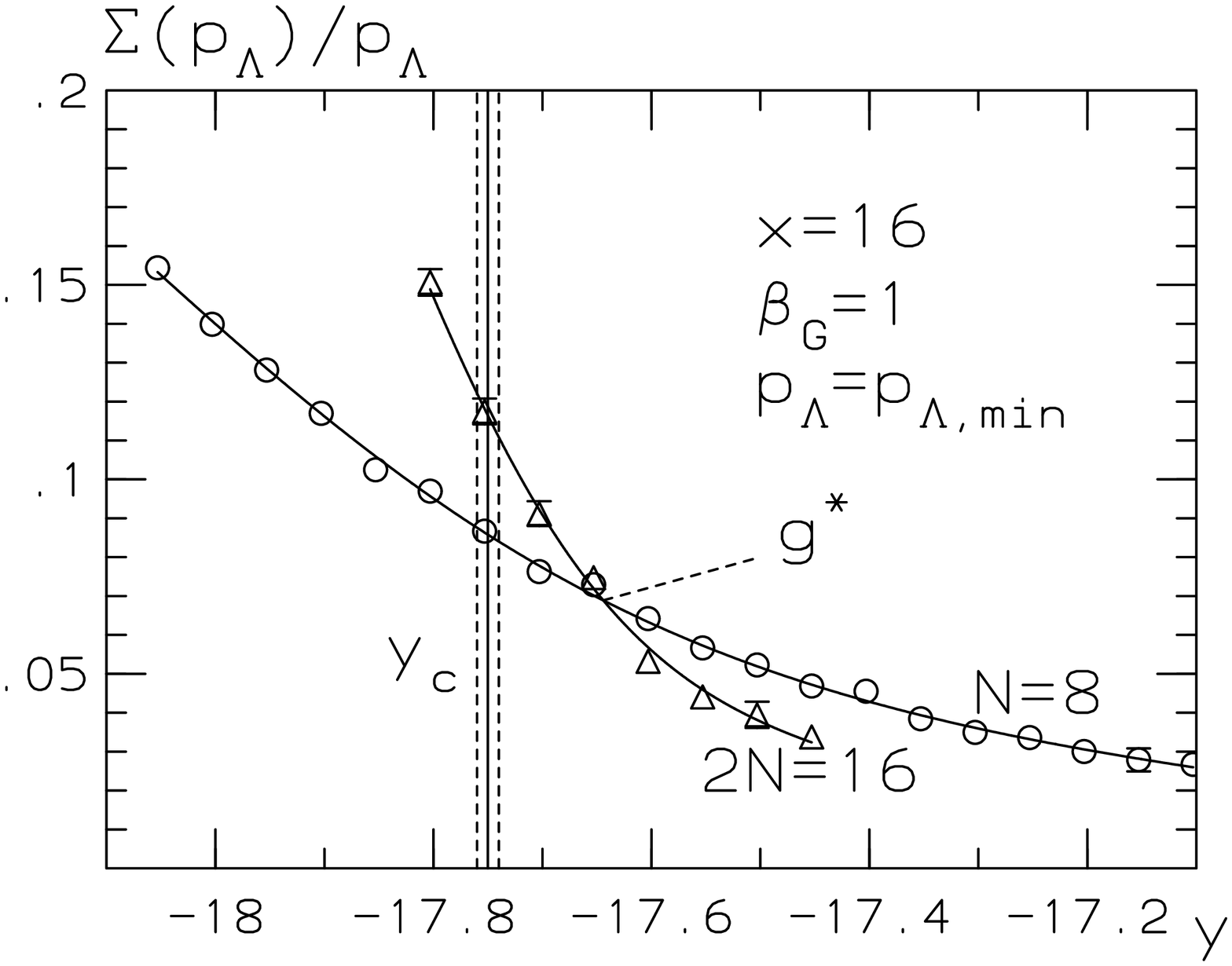,angle=270,width=7.5cm}
\psfig{file=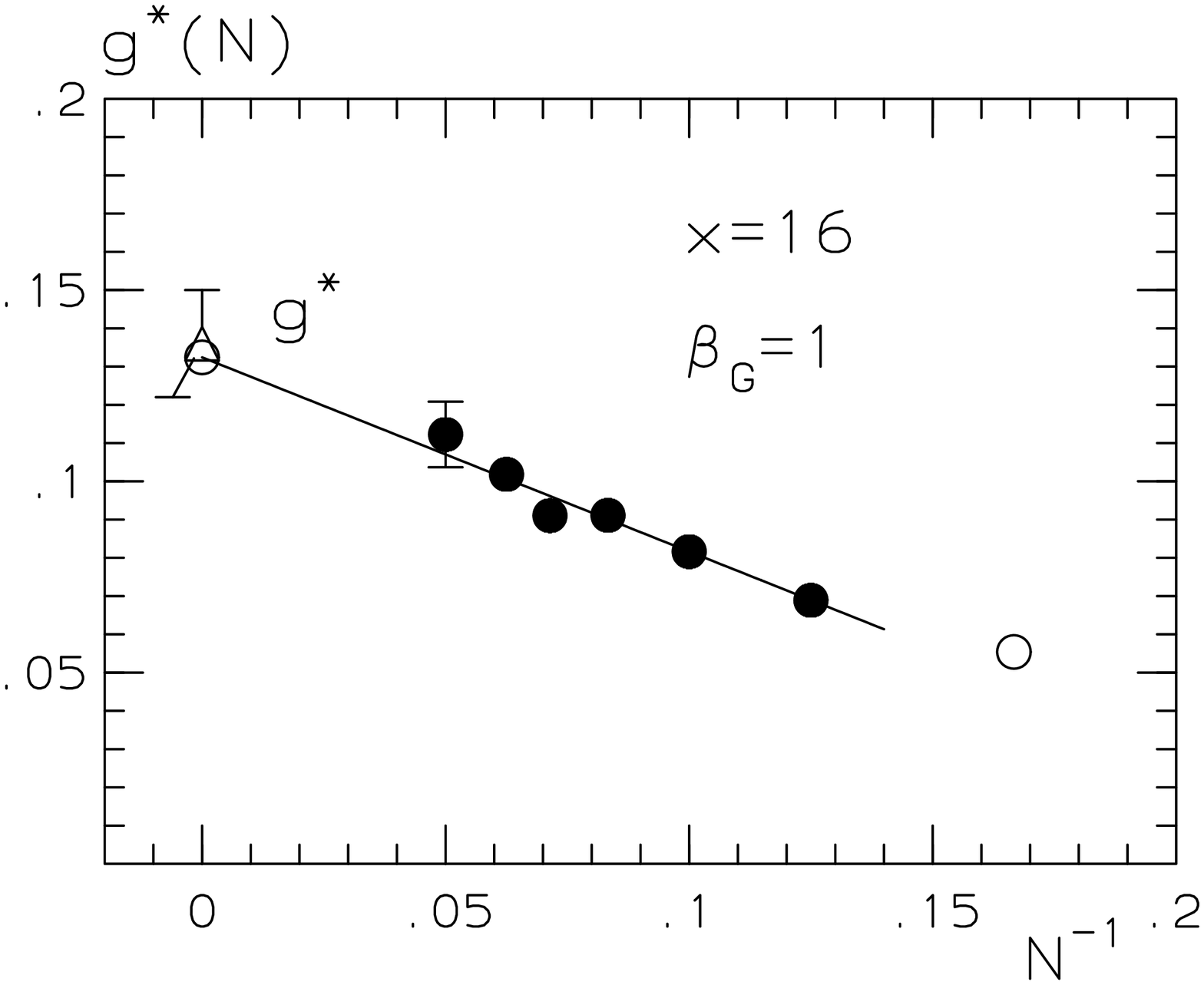,angle=270,width=7.5cm}
}

\caption[a]{ 
            Left: Determination of the fixed point value of the 
            photon self-energy from the function 
            $\Sigma(\pmin)/|\pmin|$, by a comparison
            of two volumes. 
            Right: An extrapolation to infinite volume and a comparison 
            with the determination of $g^*$ in \fig\ref{fig:eta}
            (open triangle).
           }

\la{fig:matching}
\end{figure}

\begin{figure}[t]

\centerline{
\psfig{file=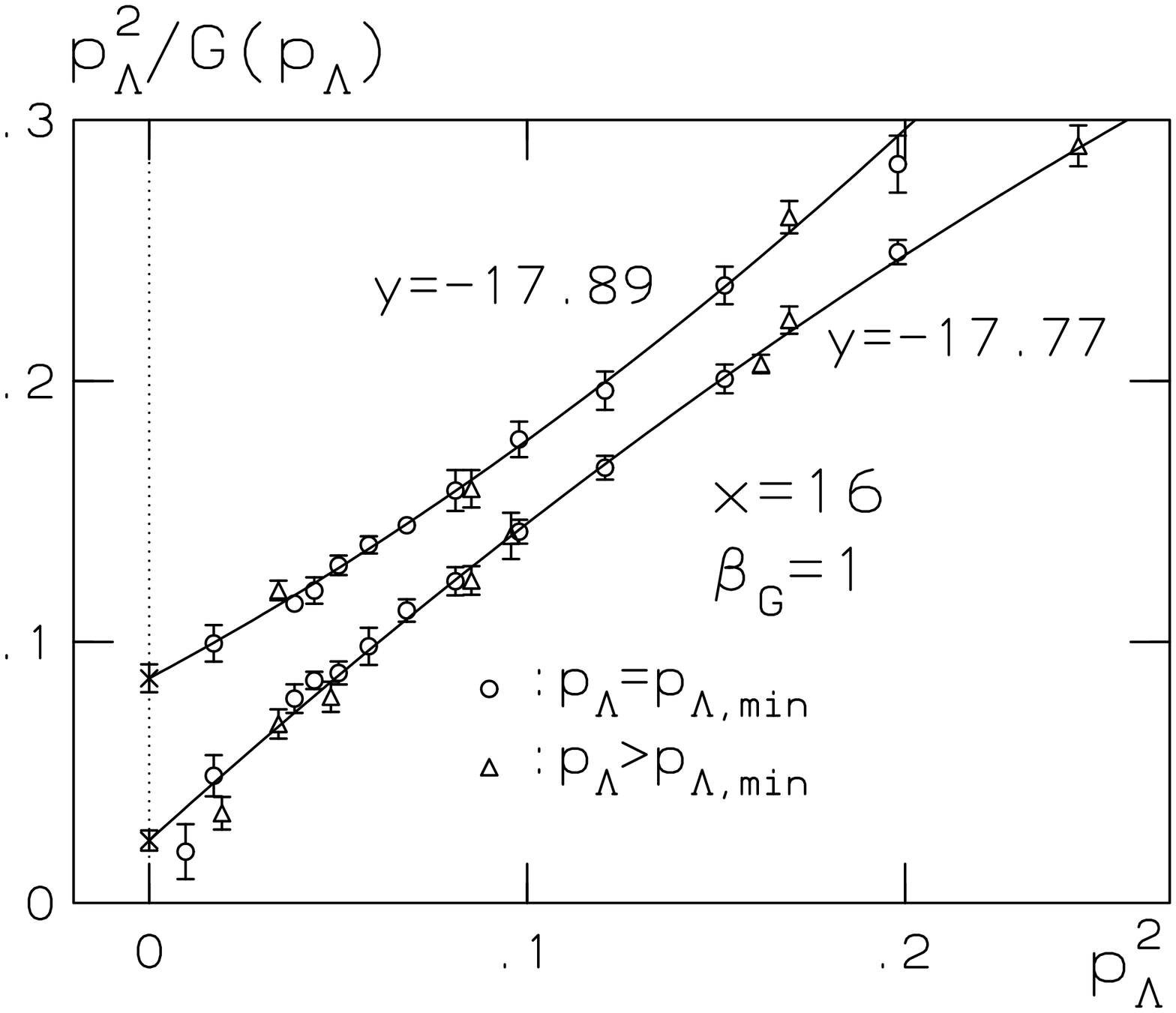,angle=270,width=7.5cm}
\psfig{file=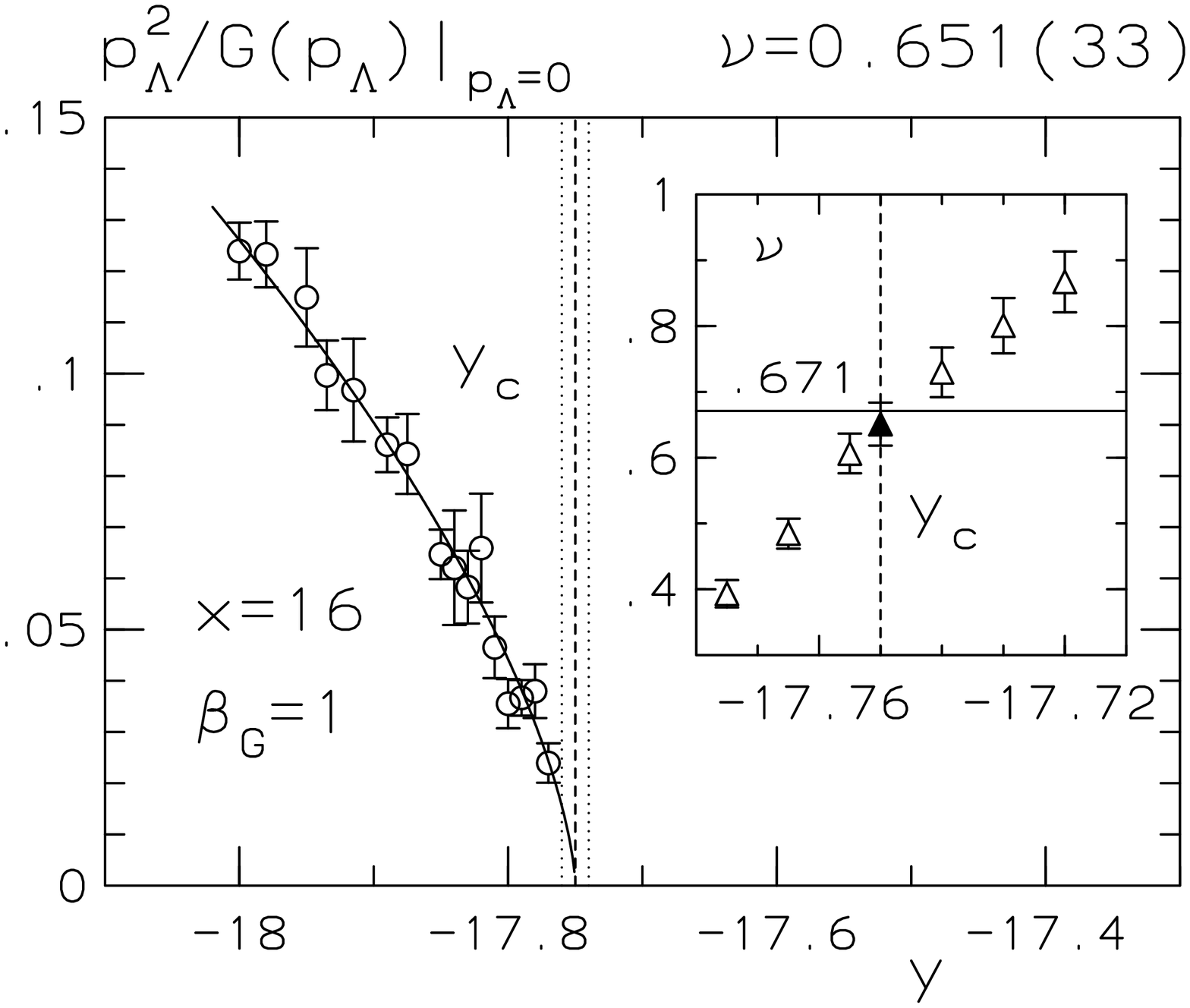,angle=270,width=7.5cm}
}

\caption[a]{ 
            Left: Infinite volume extrapolations for
            $\vec{p}^2 G^{-1}(\vec{p})$ at two values
            of $y$ in the broken symmetry phase. 
            Circles denote momenta $|\pmin|=2\pi/L$
            at different volumes; 
            triangles denote larger momenta.
            Right: The scaling behavior for
            the infinite volume extrapolated
            $\vec{p}^2 G^{-1}(\vec{p})$.
            The inlay shows the dependence of the critical exponent
            on the value of $y_c$.
           }

\la{fig:mA}
\end{figure}

\begin{figure}[t]

\centerline{
\psfig{file=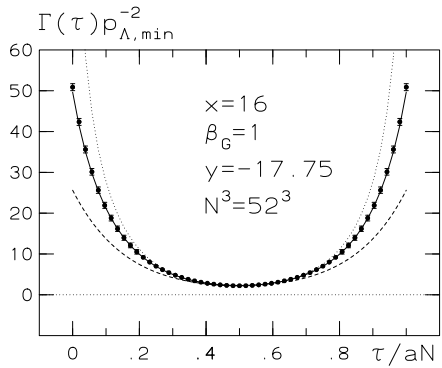,angle=270,width=7.5cm} 
}

\caption[a]{ 
            The photon correlation function at the critical point, with 
            transverse momentum $\pmin$, compared 
            with a Fourier transform of
            $1/[p^2 + A |p|]$ (solid; cf.\ \eqs\nr{Gstr2}, \nr{Gstructure}), 
            as well as with Fourier transforms of
            $1/p^2$ (dashed), $1/|p|$ (dotted),
            in each case with an overall constant chosen so that data points
            are matched at $\tau / a N = 0.5$.
           }

\la{fig:vector}
\end{figure}

\begin{figure}[t]

\psfig{file=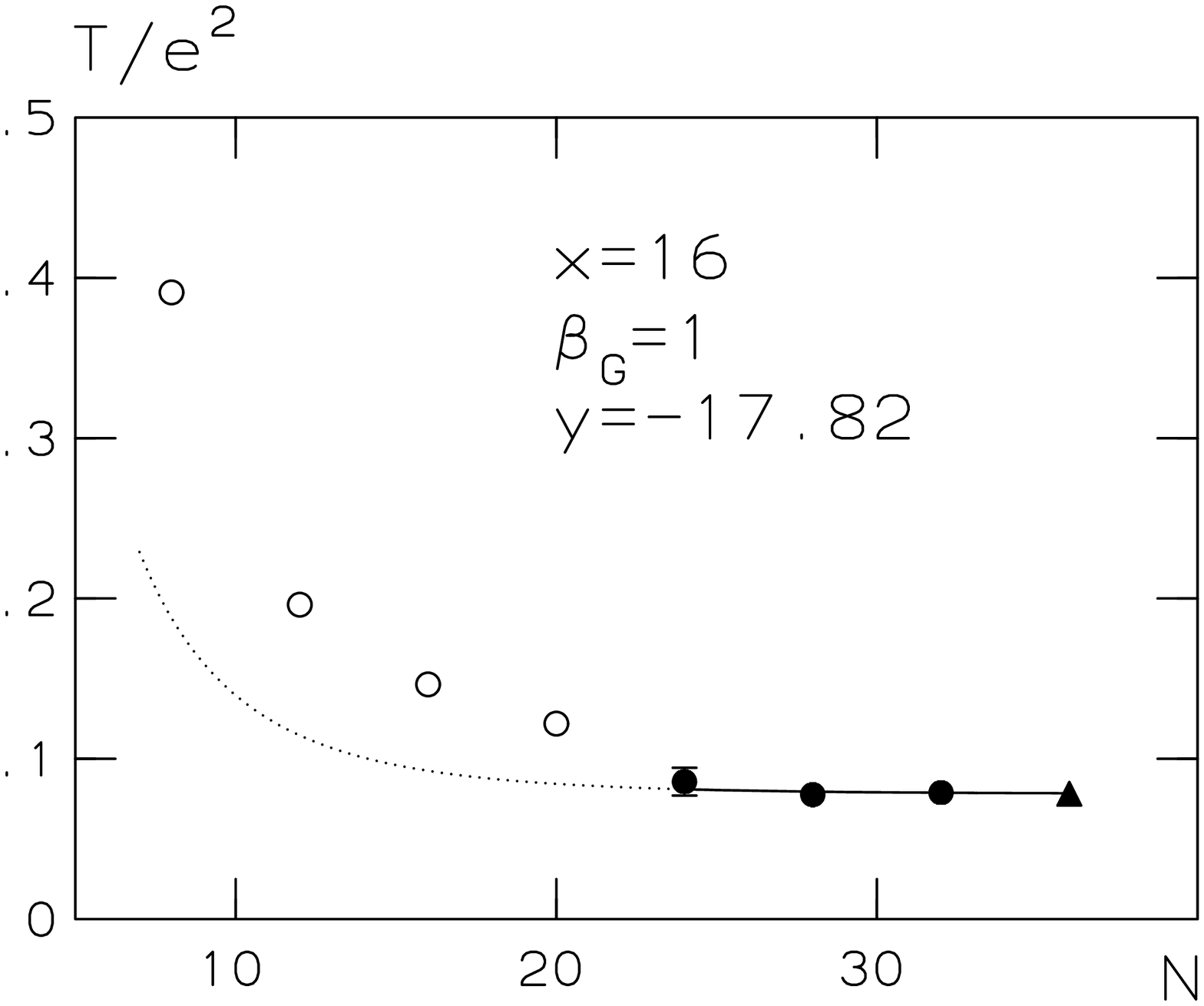,angle=270,width=7.5cm}
\psfig{file=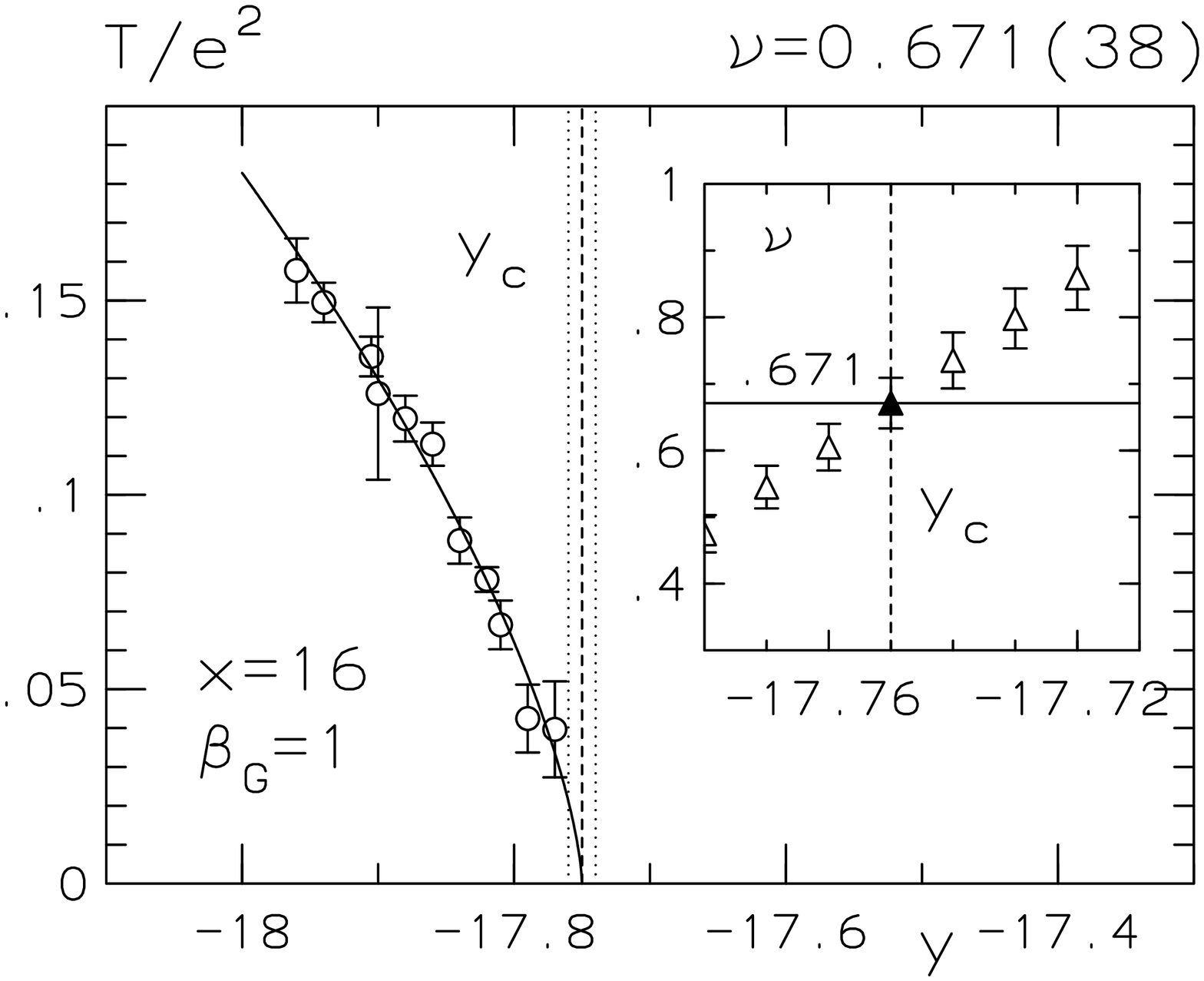,angle=270,width=7.5cm}

\caption[a]{ 
            Left: 
            An example of an infinite volume extrapolation (triangle)
            for the tension $T/e^2$
            in the broken symmetry phase. The ansatz is as explained
            in Ref.~\cite{manyvortex}, and is only applicable in large 
            volumes (solid curve).
            Right: The scaling behavior for
            the infinite volume extrapolation.
            The inlay shows the dependence of the critical exponent
            on the value of $y_c$.
           }

\la{fig:tension}
\end{figure}

%
\subsection{Discretised action}

We discretise the action in~\eq\nr{Lsqed} in a standard way, by replacing
\ba
 F_{kl}(\vec{x}) & \to & \frac{1}{ea^2} \Bigl[ 
 \alpha_k(\vec{x})+\alpha_l(\vec{x}+a \hat k)- 
 \alpha_k(\vec{x}+a \hat l)-\alpha_l(\vec{x})   \Bigr]
 \equiv \frac{1}{ea^2} \alpha_{kl}(\vec{x}) 
 \;, \la{Fkl} \\
 D_k \phi & \to & \frac{1}{a} \Bigl[ 
 \exp(i\alpha_k(\vec{x})) \phi(\vec{x} + a \hat k) - \phi(\vec{x})
 \Bigr] \;, \\
 \int {\rm d}^3 \vec{x} & \to & \sum_{\vec{x}} a^3 \;, 
\ea 
where $a$ is the lattice spacing,
$\hat k, \hat l$ are unit vectors, 
and $\alpha_i(\vec{x}) = a e A_i(\vec{x})$.\footnote{%
  Note that in~\eq\nr{Fkl} we use the non-compact formulation 
  for the U(1) gauge field. We do this to avoid topological 
  artifacts, monopoles, which appear
  in the so-called compact formulation~\cite{pol}, 
  unless $e^2a \ll 1$, and would make it difficult to reach 
  large volumes in physical units, i.e. $V \gg 1/e^6$~\cite{u1big}.} 
The bare mass parameter is written in a form which guarantees that the 
$\msbar$ renormalised mass parameter $m^2(\bmu)$, evaluated at the 
scale $\bmu = e^2$, remains finite in the continuum 
limit~\cite{contlatt}:
\ba
 m^2 & \equiv & m^2(e^2) - (e^2 + 2 \lambda)
 \frac{3.175911535625}{2\pi a} 
 \nn 
 & & 
 -\frac{1}{16\pi^2}\biggl[ 
 (-4 e^4 + 8 \lambda e^2 - 8 \lambda^2)
 \biggl(
 \ln\frac{6}{a e^2} + 0.08849 
 \biggr) - 1.1068 e^4 + 4.6358 \lambda e^2
 \biggr] \;. ~~~~~
\ea
The couplings $e^2,\lambda$, on the other hand, 
do not require renormalisation
in three dimensions
(for ${\cal O}(a)$-corrections, see Ref.~\cite{moore_a}). 
Once we also rewrite $\phi \equiv e \hat \phi$, the action becomes 
dimensionless, parameterised only by
\be
 x \equiv \frac{\lambda}{e^2}\;, \quad
 y \equiv \frac{m^2(e^2)}{e^4}\;, \quad
 \beta_G \equiv \frac{1}{e^2 a} \;.
\ee
The continuum limit corresponds to $\beta_G \to \infty$.
Instead of~\eq\nr{taudef}, we can now write
\be
 \tau = y - y_c 
 \;.
\ee 

The fields that are updated are the real variables
$\alpha_k(\vec{x})$ and the complex ones $\hat \phi(\vec{x})$.
For details concerning the update algorithm employed in this work, 
we refer to Refs.~\cite{tension,manyvortex}.
Once the system is put on the lattice, one also has to impose 
boundary conditions. In the following paragraphs
we employ periodic boundary conditions for 
all the fields. This implies that the
net winding (or flux) through the lattice in any configuration, 
$w \equiv e \int \! {\rm d}^2\, \vec{s} \cdot \vec{B}$, 
vanishes, rather than fluctuates as it in principle 
should in the canonical ensemble of \eq\nr{ZsqedH}.  
At the critical point this choice 
affects some quantities, like amplitude ratios, 
but it is not expected to
affect the critical exponents on which we concentrate here.

The observables are discretised in a straightforward way. 
In particular, denoting again $\vec{p} = 2 \pi n \hat 1/L$, the 
quantity $G(\vec{p})$ in \eq\nr{Gdef} is measured as 
\be
 G(\vec{p}) \equiv 
 \beta_G 
 \sum_\vec{x} e^{i \vec{p}\cdot \vec{x}}
 \langle \alpha_{12}(0) \alpha_{12}(\vec{x}) \rangle
 \;. \la{Gplatt}
\ee
In the figures
we always plot the momentum in the form in which it appears
in discrete space, 
\be
 p_\Lambda \equiv \frac{2}{a} \sin \frac{a |\vec{p}|}{2}
 \;,  \quad
 p_{\Lambda,\rmi{min}} \equiv \frac{2}{a} \sin \frac{a |\pmin|}{2}
 = \frac{2}{a} \sin \frac{\pi}{N} \;, 
\ee
where $\vec{p}$ was assumed to point along the $x_1$-axis, 
and the lattice
size was denoted by $L = N a$.
We often use furthermore lattice units, $a=1$.

%
\subsection{Simulation parameters}

All the simulations in this paper
have been carried out with a single lattice spacing, 
$\beta_G = 1$. The reason is that since 
we are interested in universal critical 
behaviour and are using the non-compact formulation, 
there is no need for a continuum extrapolation, 
as long as there is no phase transition in between the $\beta_G$
used and $\beta_G = \infty$; this indeed is the case. 
The scalar coupling is chosen to lie comfortably in the type II region
(i.e., beyond the Bogomolny point, $x=1/2$),
$x=16$. We have also performed some simulations at $x=2$, which
still lies in the type II region~\cite{sudbo}, 
confirming the qualitative pattern
observed at $x=16$ but with significantly less resolution. Volumes
have been chosen in the range $6^3...52^3$, and for each parameter
value we collect from $\sim 10^5$ to $\sim 2 \times 10^6$
sweeps. Different parameter values are joined together with 
Ferrenberg-Swendsen multihistogram reweighting~\cite{fs}. 

%
\subsection{Location of the critical point}

The first task is to locate the critical point, $y_c$. 
While there are many possibilities for doing this, all are
in principle equivalent in the limit $V\to\infty$, and we choose
here to use the location of the maximum of
the ``specific heat'', or the susceptibility related
to $|\hat \phi|^2$, 
\be
 \chi(|\hat \phi|^2) \equiv N^3 \Bigl[ 
 \langle (\,\overline{|\phi|^2}\, )^2 \rangle - 
 \langle  \,\overline{|\phi|^2}\, \rangle^2 \Bigr]
 \;,  
\ee
where $\overline{|\phi|^2} \equiv V^{-1} \int_x \hat\phi^* \hat \phi$.
Our data at a few representative volumes are shown in \fig\ref{fig:phi2}(left),
and an infinite volume extrapolation 
based on the positions of the susceptibility
maxima in \linebreak
\fig\ref{fig:phi2}(right). The extrapolation has been carried
out with the finite-size scaling ansatz
\be
 y_c(N=\infty) = y_c(N) +  
 c_1 \frac{1}{N^{1/\nu}} + 
 c_2 \frac{1}{N^{1/\nu + \omega}} + ...
 \;,
\ee
where the exponents have been fixed to their SFT values, 
$\nu \equiv \nu_\rmi{XY} = 0.67155(27)$, 
$\omega \equiv \omega_\rmi{XY} = 0.79(2)$~\cite{xy}. We find
\be
 y_c = -17.749(5)
 \;.
\ee
This value will be frequently referred to below,
in the form $y_c \approx -17.75$.

The critical point is of course clearly visible also in the observable
we are actually interested in, $G(\vec{p})$. In \fig\ref{fig:gs} 
we show the structure of $G(\pmin)$ (left) 
and $\pmin^2 G^{-1}(\pmin)$ (right), with 
$\pmin = 2 \pi \hat 1/L$.
Both become order parameters in the infinite volume limit $\pmin\to 0$, 
vanishing on one side of $y_c$.

%
\subsection{Symmetric phase: magnetic permeability}

The magnetic permeability, as defined in~\eq\nr{chiM}, could in principle
be determined directly from an infinite volume ($\pmin\to 0$)
extrapolation of the data in~\fig\ref{fig:gs}(left). Such an extrapolation
cannot be carried out in practice, however,
because the function approaches its infinite volume 
limit very slowly  close to the critical point.
This is illustrated in \fig\ref{fig:permn}(left) at two selected $y$-values
well inside the Coulomb phase, $y > y_c$. The corresponding extrapolated
values are shown in \fig\ref{fig:permn}(right)
(without any estimate of the systematic uncertainties introduced by 
the extrapolation), but are reliably extracted
only so far above $y_c$ that no meaningful
fit for a critical exponent can be carried out.

Fortunately, we can determine the exponent directly, by carrying
out a finite-size scaling study exactly at $y = y_c$, employing
\eq\nr{fss}. In order to measure the derivative, we construct 
explicitly the corresponding operator, 
\ba
 - \frac{{\rm d}}{{\rm d}y} G^{-1} & = &  \frac{1}{G^2}
   \frac{{\rm d}}{{\rm d}y} G 
   = - \frac{1}{G^2\beta_G^2} \Bigl[ 
   \Bigl\langle
   \sum_\vec{x} e^{i \vec{p}\cdot \vec{x}}
   \alpha_{12}(0) \alpha_{12}(\vec{x}) 
   \sum_\vec{y} \hat\phi^*(\vec{y}) \hat\phi(\vec{y})
   \Bigr\rangle 
   \nn & & \hphantom{ \frac{1}{G^2}
   \frac{{\rm d}}{{\rm d}y} G 
   = - \frac{1}{G^2\beta_G^2} \Bigl[ }
   -
   \Bigl\langle
   \sum_\vec{x} e^{i \vec{p}\cdot \vec{x}}
   \alpha_{12}(0) \alpha_{12}(\vec{x}) 
   \Bigr\rangle 
   \Bigl\langle
   \sum_\vec{y} \hat\phi^*(\vec{y}) \hat\phi(\vec{y})
   \Bigr\rangle 
   \Bigr] \;.
\ea
The data are shown in~\fig\ref{fig:permfss}(left), and a fit to the filled
circles produces a value
\be
 \nu_\chi = 0.659(26) \;.
\ee
This is well consistent with the SFT value 
$\nu_\rmi{XY} = 0.67$, as predicted by~\eq\nr{nuX}. 

%
\subsection{Transition point: anomalous dimension}

Our next task is to determine the anomalous dimension. 
In \fig\ref{fig:eta}(left), we do this directly through \eq\nr{Ap}, 
with $\Sigma(\vec{p})$ extracted from the measured $G(\vec{p})$
(\eq\nr{Gplatt}) via \eq\nr{Gstr2}.
A fit to the data at $\pmin$ produces
\be
 \eta = 0.98(6) \;,
\ee
in perfect agreement with the prediction 
following from~\eq\nr{anom_form}. 

An alternative determination of $\eta$
can be obtained by using~\eq\nr{limits}. 
In \fig\ref{fig:eta}(right), we show 
$\Sigma(\pmin) \pmin^{\eta - 2}$ at $y = y_c$, observing
that a fixed-point value in the infinite-volume 
($\pmin \to 0$) limit is only obtained with $\eta \approx 1$. We have
also estimated the fixed-point value $g^*$ of $\Sigma(\pmin)/\pmin$,  
corresponding to $A/e^2$ in the notation of~\eq\nr{limits}.\footnote{%
  The fixed-point value $g^*$, unlike critical exponents, is 
  possibly sensitive to the boundary conditions used. Let us reiterate
  that our simulations at this point 
  correspond to a vanishing winding (or flux), 
  $w=0$.
  }

The fact that the function 
$\Sigma(\pmin)/\pmin$ approaches
a fixed-point value, allows also for an alternative
determination of $g^*$. Indeed, following the standard 
procedure, we can estimate the infinite-volume limit
by comparing data for $\Sigma(\pmin)/\pmin$ at two lattice sizes, 
$N$ and $2N$, for various $N$. The procedure is illustrated
in~\fig\ref{fig:matching}(left), and produces values for 
$g^*$ (cf.\ \fig\ref{fig:matching}(right)) consistent with
but more precise than in~\fig\ref{fig:eta}(right): 
in the limit $N\to\infty$, 
a linear fit yields $g^*\approx 0.13(1)$.
This value will find applications in~\se\ref{bsp:vcl}.

%
\subsection{Broken symmetry phase: inverse vector propagator at zero momentum}

We next consider the parameter $m_\Sigma^2$, defined in~\eq\nr{mA}.
The infinite-volume extrapolation of $\pmin^2 G^{-1}(\pmin)$
is illustrated in~\fig\ref{fig:mA}(left), and a fit
as a function of $y$ is shown
in \fig\ref{fig:mA}(right). We find
\be
 \gamma_\Sigma = 0.651(33) \;,
\ee
in perfect agreement with~\eq\nr{gammaA}.

%
\subsection{Broken symmetry phase: vector correlation length}
\la{bsp:vcl}

Let us then discuss the
vector correlation length. As reported earlier on~\cite{prb,procs}, 
direct measurements of the vector correlation function struggle to 
show critical behaviour according to the exponent $\nu_\rmi{XY}$
as predicted by \eq\nr{mV}, 
producing rather the exponent $\nu_\rmi{XY}/2$. In the light of the
data presented in this paper, however, such a behaviour is well
understandable. Indeed, close to the critical point
the form of the vector propagator is as shown in~\eqs\nr{Gstr2}, 
\nr{Gstructure}, with $\eta = 1$. We have just determined that 
$A \approx g^* e^2 \approx 0.13 e^2$. This means that 
in order for true critical behaviour to show up 
in infinite volume, we would need to require
\be
 m_V^2 \ll 0.13 e^2 m_V \;, ~~~ \mbox{or} ~~
 m_ V \ll 0.13 e^2 \;.
\ee
This is the case, however, only extremely close to the transition point
(cf.\ Fig.~2 in Ref.~\cite{procs}).
Moreover, even when $m_V \approx 0$, finite-volume corrections
to the asymptotic form of the photon correlator are important, unless
\be
 \pmin^2 \ll 0.13 e^2 \pmin \;, 
\ee
which at $\beta_G = 1$ transforms to 
\be
 N \gg \frac{2\pi}{0.13} \approx 48 
 \;. \la{volume}
\ee
It is not easy to satisfy this inequality in practice, however.
Away from the transition point $m_V \approx 0$, or in 
volumes smaller than \eq\nr{volume}, 
on the other hand, the free term $\vec{p}^2$ will 
dominate the inverse vector propagator in~\eq\nr{Gstr2}, 
and solving for the pole position with the free part
alone, we recover the exponent
\be
 \nu_V^\rmi{(effective)}  \approx \fr12 \gamma_\Sigma = 
 \fr12 \nu_\rmi{XY} \;.
\ee
For completeness, the highly non-trivial form of the vector correlation
function is illustrated in~\fig\ref{fig:vector} at $y = y_c$. It is seen
clearly how the free term $\pmin^2$ and the linear term in
$\Sigma(\pmin)$ are both needed in order
to reproduce the data points, even at 
volumes as large as $52^3$.

If the mapping between the parameters of SQED and SFT were known, 
one would not need to rely on the asymptotic critical behaviour to 
demonstrate the duality between the two theories, but one could 
compare the data directly, staying away from the transition point 
and keeping the volume finite. Without the mapping
such a comparison is not available on a quantitative level, 
but we may still note that the data obtained for the
vector correlation length in the ``frozen superconductor'' model, where
the duality is exact, show a behaviour very similar to what 
we have argued for here and observed in Refs.~\cite{prb,procs}, i.e.\ 
$\nu_\rmi{XY}/2 \lsim \nu_V^\rmi{effective} < \nu_\rmi{XY}$, 
at least apart from the extreme vicinity of the transition point~\cite{fzs}. 
This provides further  qualitative support for the duality conjecture, through 
similar corrections to asymptotic scaling in both theories.

%
\subsection{Broken symmetry phase: vortex tension}
\la{bsp:vt}

We finally consider the vortex tension, $T$.\footnote{%
 We use the same notation for the vortex tension and the temperature, 
 but there should be no danger of confusion, since the two are never
 discussed in the same context. 
 } 
Unlike for the observables
so far, its determination requires that we depart from strictly periodic 
boundary conditions for all the fields. In fact, the Monte Carlo 
determination of $T$ proceeds via a computation of the free
energy difference 
$\ln\mathcal{Z}_\rmi{SQED}(w=0) - \ln\mathcal{Z}_\rmi{SQED}(w=2\pi)$, 
where the winding number 
$w = e \int \! {\rm d}^2 \vec{s}\cdot \vec{B} = 2\pi$ 
corresponds to the presence of a single vortex through the lattice. 
Parametrizing $w=2\pi z$,  we can write
$\ln\mathcal{Z}_\rmi{SQED}(w=2\pi)=
\ln\mathcal{Z}_\rmi{SQED}(0)+\int_{0}^{1} \! {\rm d}z 
\langle W(z) \rangle$, where $\langle W(z) \rangle$ represents 
a specific expectation value, which can be measured by Monte Carlo 
methods. For more details, we refer the reader
to our earlier work~\cite{tension,manyvortex}.

In the earlier work mentioned,
the integral $\int_{0}^{1} \! {\rm d}z \langle W(z) \rangle$
was approximated by a sum over
independent and consecutive Monte Carlo 
measurements at 
a discrete set $z_i$, $i=1,...,n_z$, where the parameter 
$n_z$ had typically values $n_z=11,...,21$.
For the current work, we have implemented ``parallel tempering''
in the parameters $z_i$, along the lines of Ref.~\cite{mp}, 
in an attempt to sample intermediate configurations with 
non-integer winding numbers $0 < z < 1$ more efficiently. Using 
again $n_z=21$, we find a sizable error reduction, as compared
with our earlier approach. This is illustrated
by the single data point with large error bars 
in \fig\ref{fig:tension}(right), at $y\approx-17.9$, obtained with 
the unimproved algorithm.  
We have not investigated the precise mechanism behind
this algorithmic improvement in great detail, however.

In \fig\ref{fig:tension}(left)
we show an example of the infinite volume extrapolation of 
the tension in the broken phase, and in \fig\ref{fig:tension}(right)
a fit to the extrapolated values. The fit yields
\be
 \nu_{T} = 0.671(38) \;, 
\ee
in perfect agreement with \eq\nr{nuT}.

%
\section{Macroscopic external magnetic fields}
\la{se:Bext}

Up to this point, we have mostly considered the case that 
the magnetic field is set to zero, $H_i = 0$, after taking
derivatives of the partition function (cf.\ \eq\nr{jjrelxy}); 
the main exception was the discussion of the vortex tension, leading
to~\eq\nr{nuT} and continued in~\se\ref{bsp:vt}. 
In this section, we wish to make some further
qualitative remarks on the case that the external magnetic 
field is not set to zero but kept finite.  

To begin with, let us reiterate the duality relation for 
this situation, 
in rather explicit form. We choose again coordinates such that the 
constant magnetic field is pointing in the third direction, 
$H_3 \equiv H$, and consider the system to live in a finite 
box (or hypertorus), now of extent $L$ in the 
third direction. Then $\mathcal{Z}_\rmi{SQED}[H]$ represents the 
{\em classical} Euclidean partition function for {\em 3-dimensional} 
SQED, with the external parameters $H,L$. At the 
same time, the corresponding $\mathcal{Z}_\rmi{SFT}[H]$, 
\eqs\nr{Lsft}, \nr{Zsft}, is just the imaginary time 
{\em quantum} partition function for a {\em 2-dimensional}
SFT, in the presence of a finite 
chemical potential $\mu = \tilde e H$ (written in a relativistic form)
and a finite temperature $T$
(fixed by the extent of the imaginary time direction). 
Thus, the relations between $H,L$
on one side and $\mu,T$ on the other, read 
\be
 H = \frac{e}{2\pi} \mu \;, \quad
 L = \frac{\hbar c}{k_B T} \;. \la{dualBext}
\ee
Taking the box cubic and sending $L \to \infty$, corresponds to 
computing the quantum partition function for the 2-dimensional SFT 
at zero temperature. While the relations in~\eq\nr{dualBext} 
have been expressed in a canonical ensemble with respect to $H,\mu$, 
it should be clear that the correspondence remains true also in 
a microcanonical ensemble (i.e., fixed magnetic flux in SQED / fixed 
particle number in SFT), with a proper choice of boundary conditions. 

\paragraph{Implications for the type I region.} 

In the type I region of SQED, 
vortices attract each other. Therefore, if we are in the broken
symmetry phase and force a magnetic
flux through the system, the vortices form a flux tube that 
penetrates through the Meissner phase. What is the analogue for 
this in SFT? The type I region corresponds to $\tilde \lambda < 0$
(and the presence of further stabilising terms), and as is well known, 
in this case the system admits non-topological soliton 
solutions~\cite{fls,lp}, or droplets of Bose liquid, 
to be intuitively thought of as non-dispersive bound states 
of particles held together by the attractive interaction.
We thus observe a perfect analogy, with the role of flux quanta in SQED
played by the attractive particles in SFT. Note that even 
though SFT appears in a 
(2+1 dimensional imaginary time) 
relativistic form in~\eqs\nr{Lsft}, \nr{Zsft}, 
the same solutions appear there
even if we approach the non-relativistic 
limit~\cite{lp,le}, a situation also familiar from 
actual experiments with soliton-like structures in
superfluid Helium (see, e.g., Ref.~\cite{gv}) 
and atomic Bose-Einstein condensates (see, e.g., Ref.~\cite{exp_bec}).

\paragraph{Implications for the type II region.} 

In the type II region of SQED, on the other hand,
vortices repel each other. Therefore, if we place the system in a
magnetic field, they form (at least on the mean field level)
an Abrikosov vortex lattice. It has 
been a long-standing issue to study whether, within SQED, 
the vortex lattice indeed exists in 
a strict sense for low enough magnetic fields, and then melts to a vortex
liquid possibly through a first order transition, as observed
experimentally~\cite{zeldov}, or always
appears in a liquid state due to fluctuations, 
so that the transition observed
in experiments would be a manifestation of some physics 
beyond the pure SQED (see, e.g., Ref.~\cite{km}). 
We now see that on the side of SFT, this 
corresponds to whether a 2-dimensional
dilute system of atoms with repulsive
interactions, forms a lattice, or melts due to quantum and thermal
fluctuations. In principle, 
this issue could be studied with SFT, 
more easily than directly with SQED as attempted in 
Ref.~\cite{manyvortex}, since the system has fewer dynamical length scales.
In practice, though, the inclusion of a chemical potential in SFT makes
the action complex for a generic configuration 
(this is just the generic ``sign problem'' for $\mu\neq 0$), 
rendering importance sampling ineffective.

%
\section{Conclusions}
\la{se:concl}

In this paper, we have used numerical lattice Monte Carlo simulations
to measure a number of critical exponents in three-dimensional scalar
electrodynamics (SQED) in the type II region, or $\lambda/e^2 \gsim 1$. 
We have also reiterated which
exponents of the dual scalar field theory (SFT) 
they should correspond to. Our measurements agree 
well with the SFT values, known to high accuracy from
previous studies of the XY model. This provides strong 
``empirical'' evidence for the existence 
of the conjectured duality between the two theories.

We have also elaborated on the qualitative implications of the duality
for the case of macroscopic external magnetic fields, 
both in the type I and in the type II regions. Even though we have
no new simulations to report in this regime, we find it remarkable that
for all the known phenomena observed in SQED somewhat below
the phase transition temperature, one indeed finds a perfect counterpart 
in 2-dimensional quantum SFT, but at a different temperature,
as determined by the physical geometry of the SQED sample. 

The duality predicts that the only important degrees of freedom near
the transition point in SQED are vortex lines, and the quantities
we have measured are sensitive to their properties. The agreement with
the dual theory therefore also demonstrates
that the methods we have used in these simulations give an accurate 
description of the dynamics of topological defects. Similar methods
can also be used in more complicated theories, e.g., in the case of
magnetic monopoles in non-Abelian theories~\cite{Davis:2000kv}.
Therefore it should be possible to use these techniques 
to carry out similar studies in those systems.

In the past, the study of dualities has often been restricted to
spin models, low dimensions, or supersymmetric theories. The reason for
this has mostly been that only in these cases has one been
able to carry out controllable analytic calculations to test and 
make use of the dualities. Our results show how
the study of dualities can in principle
be extended towards more realistic systems, 
by invoking numerical techniques to corroborate analytic arguments
based on very general principles only, such as symmetries. The same
numerical techniques could also be used to determine 
``experimentally'' the mappings between the parameter sets.
It should be mentioned that many other types of numerical 
avenues towards duality have, of course, been explored and extensively
tested in the case of 4-dimensional Yang-Mills theory
(see, e.g., Ref.~\cite{qcd} and references therein). 

To conclude, let us finally 
recall that apart from the theoretical 
issues on which we have concentrated in this paper, the critical
properties of three-dimensional SQED have also physical significance
of their own, given that this theory
is the effective theory of superconductivity~\cite{kleinert}
and of liquid crystals~\cite{lc} in certain regimes, as well as of 
four-dimensional scalar electrodynamics
at high temperatures~\cite{perturbative,u1big,joa}.
Particularly in the first two of 
these cases, the substantial corrections to scaling that
we have observed for the photon correlation
length (also called the penetration depth), 
may also have phenomenological significance~\cite{exp}, given that
it is difficult to probe very precisely the extreme vicinity of the 
transition point and that experimental samples 
are necessarily fairly restricted in size. 

%
\section*{Acknowledgements}

We acknowledge useful discussions, over the years, 
with P.~de Forcrand, 
H.~Kleinert, \linebreak A.~Kovner, I.D.~Lawrie, D.~Litim,  
F.S.~Nogueira, R.D.~Pisarski, A.M.J.~Schakel, A. Sudb{\o}, B.~Svetitsky, 
and G.E.~Volovik.
This work was partly supported by the RTN network {\em Supersymmetry}
{\em and} 
{\em the} {\em Early Universe}, EU contract no.\ HPRN-CT-2000-00152, 
by the ESF COSLAB programme,
and by the Academy of Finland, contracts no.\ 77744 and 80170.  
A.R. was supported by Churchill College, Cambridge.


\appendix
\renewcommand{\thesection}{Appendix~\Alph{section}}
\renewcommand{\thesubsection}{\Alph{section}.\arabic{subsection}}
\renewcommand{\theequation}{\Alph{section}.\arabic{equation}}




\begin{thebibliography}{99}

\bibitem{Kramers:1941kn}
H.A.~Kramers and G.H.~Wannier,
Phys.\ Rev.\  {60} (1941) 252.

\bibitem{gauge}
R.\ Balian, J.M.\ Drouffe and C. Itzykson,
Phys.\ Rev.\ D {11} (1975) 2098;
%
C.P.\ Korthals Altes, 
Nucl.\ Phys.\ B {142} (1978) 315;
%
T.~Yoneya,
Nucl.\ Phys.\ B {144} (1978) 195.

\bibitem{peskin}
M.E.~Peskin,
Annals Phys.\  {113} (1978) 122.

\bibitem{hk}
S.~Hands and J.B.~Kogut,
Nucl.\ Phys.\ B {462} (1996) 291
[hep-lat/9509072].

\bibitem{Coleman:bu}
S.R.~Coleman,
Phys.\ Rev.\ D {11} (1975) 2088.

\bibitem{Mandelstam:1975hb}
S.~Mandelstam,
Phys.\ Rev.\ D {11} (1975) 3026.

\bibitem{sw}
N.~Seiberg and E.~Witten,
Nucl.\ Phys.\ B {426} (1994) 19; 
{\em ibid.}\ B {430} (1994) 485 (E)
[hep-th/9407087];
%
{\em ibid.}\ B {431} (1994) 484
[hep-th/9408099].

\bibitem{Montonen:1977sn}
C.~Montonen and D.I.~Olive,
Phys.\ Lett.\ B {72} (1977) 117.

\bibitem{Osborn:tq}
H.~Osborn,
Phys.\ Lett.\ B {83} (1979) 321.

\bibitem{Maldacena:1997re}
J.M.~Maldacena,
Adv.\ Theor.\ Math.\ Phys.\  {2} (1998) 231
[hep-th/9711200].

\bibitem{qcd}
G.~Ripka,
{\em Dual superconductor models of color confinement},
hep-ph/0310102.



\bibitem{bmk}
T. Banks, R. Myerson and J.B. Kogut, 
Nucl.\ Phys.\ B 129 (1977) 493.

\bibitem{ts}
P.R. Thomas and M. Stone, 
Nucl.\ Phys.\ B 144 (1978) 513.

\bibitem{rs}
R. Savit, 
Phys.\ Rev.\ B 17 (1978) 1340. 

\bibitem{dh} 
C. Dasgupta and B.I. Halperin, 
Phys.\ Rev.\ Lett.\ 47 (1981) 1556.

\bibitem{kleinert2} 
H. Kleinert, 
Lett.\ Nuovo Cim.\ {35} (1982) 405;
%
M. Kiometzis, H. Kleinert and A.M.J. Schakel,
%
Fortsch.\ Phys.\ {43} (1995) 697
[cond-mat/9508142].

\bibitem{kleinert}
H. Kleinert, {\it Gauge Fields in Condensed Matter}, vol.\ 1
(World Scientific, Singapore, 1989).

\bibitem{kovner} 
A. Kovner, B. Rosenstein and D. Eliezer, 
Nucl.\ Phys.\ B {350} (1991) 325;
%
A. Kovner, P. Kurzepa and B. Rosenstein, 
Mod.\ Phys.\ Lett.\ A {8} (1993) 1343; 
{\em ibid.} A {8} (1993) 2527 (E)
[hep-th/9303144].

\bibitem{ifh}
I.F. Herbut,
J.\ Phys.\ A: Math.\ Gen.\ 30 (1997) 423 [cond-mat/9610052].

\bibitem{son}
D.T.~Son,
JHEP {02} (2002) 023
[hep-ph/0201135].



\bibitem{xy}
R.~Guida and J.~Zinn-Justin,
J.\ Phys.\ A:\ Math.\ Gen.\ 31 (1998) 8103
[cond-mat/9803240];
%
M.~Hasenbusch and T.~T\"or\"ok,
J.\ Phys.\ A:\ Math.\ Gen.\ 32 (1999) 6361
[cond-mat/9904408];
%
M.~Campostrini, M.~Hasenbusch, A.~Pelissetto, P.~Rossi and E.~Vicari,
Phys.\ Rev.\ B 63 (2001) 214503
[cond-mat/0010360].

\bibitem{tension}
K.~Kajantie, M.~Laine, T.~Neuhaus, J.~Peisa, A.~Rajantie and K.~Rummukainen,
Nucl.\ Phys.\ B {546} (1999) 351
[hep-ph/9809334].

\bibitem{manyvortex}
K.~Kajantie, M.~Laine, T.~Neuhaus, A.~Rajantie and K.~Rummukainen,
Nucl.\ Phys.\ B {559} (1999) 395
[hep-lat/9906028].

\bibitem{ot}
P.~Olsson and S.~Teitel,
Phys.\ Rev.\ Lett.\  {80} (1998) 1964
[cond-mat/9710200].

\bibitem{fzs}
T.~Neuhaus, A.~Rajantie and K.~Rummukainen,
Phys.\ Rev.\ B {67} (2003) 014525
[cond-mat/0205523].

\bibitem{hlm}
B.I. Halperin, T.C. Lubensky and S.-K. Ma, 
Phys.\ Rev.\ Lett.\ 32 (1974) 292.

\bibitem{fh}
R. Folk and Yu. Holovatch, 
J.\ Phys.\ A 29 (1996) 3409; 
%
cond-mat/9807421. 

\bibitem{bfllw} 
B. Bergerhoff, F. Freire, D.F. Litim, S. Lola and C. Wetterich, 
Phys.\ Rev.\ B {53} (1996) 5734
[hep-ph/9503334].

\bibitem{ht}
I. Herbut and Z. Te{\v s}anovi{\' c},
Phys.\ Rev.\ Lett.\ {76} (1996) 4588 [cond-mat/9605185].

\bibitem{kn}
H.~Kleinert and F.S.~Nogueira,
Nucl.\ Phys.\ B {651} (2003) 361
[cond-mat/0104573].

\bibitem{b} 
J. Bartholomew, 
Phys.\ Rev.\ B 28 (1983) 5378.

\bibitem{munehisa}
Y. Munehisa, 
Phys.\ Lett.\ B {155} (1985) 159.

\bibitem{prb}
K.~Kajantie, M.~Karjalainen, M.~Laine and J.~Peisa,
Phys.\ Rev.\ B {57} (1998) 3011
[cond-mat/9704056].

\bibitem{sudbo}
A.K. Nguyen and A. Sudb{\o},
Phys.\ Rev.\ B 60 (1999) 15307 
[cond-mat/9907385];
%
S. Mo, J. Hove and A. Sudb{\o},
Phys.\ Rev.\ B 65 (2002) 104501
[cond-mat/0109260]. 

\bibitem{procs}
K.~Kajantie, M.~Laine, T.~Neuhaus, A.~Rajantie and K.~Rummukainen,
Nucl.\ Phys.\ B (Proc.\ Suppl.)\  {106} (2002) 959
[hep-lat/0110062].

\bibitem{m-r} 
J.~March-Russell,
Phys.\ Lett.\ B {296} (1992) 364
[hep-ph/9208215].

\bibitem{idl}
I.D.~Lawrie, 
Nucl.\ Phys.\ B {200} (1982) 1;
%
I.D.~Lawrie and C.~Athorne,
J.\ Phys.\ A {16} (1983) L587.

\bibitem{weinberg}
S.~Weinberg,
Physica A {96} (1979) 327.

\bibitem{gl}
J.~Gasser and H.~Leutwyler,
Annals Phys.\  {158} (1984) 142;
%
Nucl.\ Phys.\ B {250} (1985) 465.

\bibitem{kapusta}
J.I.~Kapusta, 
{\em Finite Temperature Field Theory}
(Cambridge University Press, Cambridge, 1989).

\bibitem{fbj}
B.D. Josephson, 
Phys.\ Lett.\ 21 (1966) 608; 
%
M.E. Fisher, M.N. Barber and D. Jasnow, 
Phys.\ Rev.\ A 8 (1973) 1111.

\bibitem{ap}
T. Appelquist and R.D. Pisarski,
Phys.\ Rev.\ D 23 (1981) 2305.

\bibitem{hs}
J. Hove and A. Sudb{\o},
Phys.\ Rev.\ Lett.\ 84 (2000) 3426
[cond-mat/0002197].

\bibitem{pol}
A.M.~Polyakov,
Phys.\ Lett.\ B {59} (1975) 82;
%
Nucl.\ Phys.\ B {120} (1977) 429.


\bibitem{u1big}
K.~Kajantie, M.~Karjalainen, M.~Laine and J.~Peisa,
Nucl.\ Phys.\ B {520} (1998) 345
[hep-lat/9711048].

\bibitem{contlatt}
M.~Laine and A.~Rajantie,
Nucl.\ Phys.\ B {513} (1998) 471
[hep-lat/9705003].

\bibitem{moore_a}
G.D. Moore, 
Nucl.\ Phys.\ B {523} (1998) 569
[hep-lat/9709053].

\bibitem{fs}
A.M.~Ferrenberg and R.H.~Swendsen,
Phys.\ Rev.\ Lett.\  {63} (1989) 1195.

\bibitem{mp}
E.~Marinari and G.~Parisi,
Europhys.\ Lett.\  {19} (1992) 451
[hep-lat/9205018].

\bibitem{fls}
G.~Rosen,
J.\ Math.\ Phys.\ 9 (1968) 996;
%
R.~Friedberg, T.D.~Lee and A.~Sirlin,
Phys.\ Rev.\ D {13} (1976) 2739.

\bibitem{lp}
T.D.~Lee and Y.~Pang,
Phys.\ Rept.\  {221} (1992) 251,
%
and references therein.

\bibitem{le}
K.~Enqvist and M.~Laine,
JCAP {08} (2003) 003
[cond-mat/0304355].

\bibitem{gv}
G.E. Volovik, 
{\em The Universe in a Helium Droplet}, Chapter 3.3
(Clarendon Press, Oxford, 2003);
%
V.V. Dmitriev, V.B. Eltsov, M. Krusius, J.J. Ruohio and G.E. Volovik,
Phys.\ Rev.\ B 59 (1999) 165
[cond-mat/9805119].

\bibitem{exp_bec}
L.~Khaykovich {\em et al.}, 
Science 296 (2002) 1290
[cond-mat/0205378];
%
K.E.~Strecker {\em et al.}, 
Nature 417 (2002) 150
[cond-mat/0204532].

\bibitem{zeldov}
E. Zeldov {\em et al.}, 
Nature 375 (1995) 373; 
%
A. Schilling {\em et al.}, 
Nature 382 (1996) 791.

\bibitem{km}
A.K. Kienappel and M.A. Moore,
Phys.\ Rev.\ B {60} (1999) 6795 
[cond-mat/9809317],
and references therein.

\bibitem{Davis:2000kv}
A.C.~Davis, T.W.B.~Kibble, A.~Rajantie and H.~Shanahan,
JHEP {11} (2000) 010
[hep-lat/0009037];
A.C.~Davis, A.~Hart, T.W.B.~Kibble and A.~Rajantie,
Phys.\ Rev.\ D {65} (2002) 125008
[hep-lat/0110154].

\bibitem{lc}
P.G. de Gennes, 
Solid State Commun.\ 10 (1972) 753;
%
B.I. Halperin and T.C. Lubensky, 
{\em ibid.}\ 14 (1974) 997.

\bibitem{perturbative}
K.~Farakos, K.~Kajantie, K.~Rummukainen and M.~Shaposhnikov,
Nucl.\ Phys.\ B {425} (1994) 67
[hep-ph/9404201].

\bibitem{joa}
J.O. Andersen, 
Phys.\ Rev.\ D 59 (1999) 065015.

\bibitem{exp}
S. Kamal {\em et al.}, 
Phys.\ Rev.\ Lett.\ 73 (1994) 1845;
%
S. Kamal {\em et al.}, 
Phys.\ Rev.\ B 58 (1998) R8933; 
%
K.M. Paget, B.R. Boyce and T.R. Lemberger, 
Phys.\ Rev.\ B 59 (1999) 6545.

\end{thebibliography}
\end{document}